\documentclass[manuscript,authorversion]{acmart}

\AtBeginDocument{%
  \providecommand\BibTeX{{%
    \normalfont B\kern-0.5em{\scshape i\kern-0.25em b}\kern-0.8em\TeX}}}

\setcopyright{acmcopyright}
\copyrightyear{2024}
\acmYear{2024}
\acmDOI{XXXXXXX.XXXXXXX}

\usepackage{algorithm}
\usepackage{algpseudocode}
\usepackage{multirow,epsfig,bm,pifont}
\usepackage{mathtools, float}
\usepackage[toc,page]{appendix}
\usepackage{comment}
\usepackage{spverbatim}
\usepackage{xcolor}
\usepackage{subcaption}
\usepackage{todonotes}

\newcommand{\beq}{\begin{equation}}
\newcommand{\eeq}{\end{equation}}
\newcommand{\bea}{\begin{eqnarray}}
\newcommand{\eea}{\end{eqnarray}}

\usepackage{fancyvrb} 
\usepackage{mathtools}
\DeclarePairedDelimiter\ceil{\lceil}{\rceil}

\begin{document}
\setcopyright{licensedusgovmixed}
\acmYear{2024} \acmVolume{5} \acmNumber{4} \acmArticle{25} \acmMonth{10}\acmDOI{10.1145/3689826}
\title{Realistic Cost to Execute Practical Quantum Circuits using Direct Clifford+T Lattice Surgery Compilation}

\author{Tyler LeBlond}
\affiliation{%
  \institution{Computational Sciences and Engineering Division \\ Oak Ridge National Laboratory}
  \city{Oak Ridge}
  \country{USA}}
\email{leblondtr@ornl.gov}

\author{Christopher Dean}
\affiliation{%
  \institution{Department of Mathematics and Statistics \\ Dalhousie University}
  \city{Halifax}
  \country{Canada}}
\email{christopher.dean@dal.ca}

\author{George Watkins}
\affiliation{%
  \institution{Department of Computer Science \\ Aalto University}
  \city{Espoo}
  \country{Finland}}
\email{invio.george@gmail.com}

\author{Ryan S. Bennink}
\affiliation{%
  \institution{Computational Sciences and Engineering Division \\ Oak Ridge National Laboratory}
  \city{Oak Ridge}
  \country{USA}}
\email{benninkrs@ornl.gov}

\begin{abstract}

We report a resource estimation pipeline that explicitly compiles quantum circuits expressed using the Clifford+T gate set into a surface code lattice surgery instruction set. The cadence of magic state requests from the compiled circuit enables the optimization of magic state distillation and storage requirements in a \textit{post-hoc} analysis. To compile logical circuits into lattice surgery operations, we build upon the open-source Lattice Surgery Compiler. The revised compiler operates in two stages: the first translates logical gates into an abstract, layout-independent instruction set; the second compiles these into local lattice surgery instructions that are allocated to hardware tiles according to a specified resource layout. The second stage retains logical parallelism while avoiding resource contention in the fault-tolerant layer, aiding realism. Additionally, users can specify dedicated tiles at which magic states are replenished, enabling resource costs from the logical computation to be considered independently from magic state distillation and storage. We demonstrate the applicability of our pipeline to large, practical quantum circuits by providing resource estimates for the ground state estimation of molecules. We find that variable magic state consumption rates in real circuits can cause the resource costs of magic state storage to dominate unless production is varied to suit.

\end{abstract}

\maketitle

\section{Introduction}
\label{sec:introduction}

In quantum computing, promises of exponential speed-ups for certain classes of computationally hard problems have spurred an intense focus on research and development of systems capable of fulfilling these promises~\cite{gibney2019quantum, parker2022assessment}. It is generally understood that, due to the fragility of quantum states, it is necessary to encode logical qubit states within the joint Hilbert space of many physical qubits using what are called quantum error-correcting codes (QECC)~\cite{gottesman1997stabilizer, terhal2015quantum, campbell2017roads}. Currently, the surface code is the leading QECC due to the locality of its error checks (implying straightforward hardware implementation) and high error threshold (implying greater fault-tolerance)~\cite{kitaev2003fault, fowler2012surface, litinski2019game}. While it can be said that we are near the fault-tolerant era because there have been demonstrations of fault-tolerance by major hardware providers~\cite{ryan2021realization, abobeih2022fault, google2023suppressing, da2024demonstration}, the overhead of surface code quantum computing is large enough that leading systems may not have enough qubits to implement practical quantum algorithms fault-tolerantly for a long time. Still, there has already been much discussion and refinement in the literature regarding strategies to compile logical circuits into fault-tolerant operations on the surface code\footnote{We will call this concept \textit{fault-tolerant compilation}.}, with lattice surgery offering the leading instruction set~\cite{horsman2012surface,herr2017lattice,herr2017optimization,brown2017poking,fowler2018low,litinski2018lattice,litinski2019game,beverland2022surface,beverland2022assessing, chamberland2022universal,chamberland2022circuit}, and with some open-source compilers already having been developed~\cite{paler2020opensurgery,watkins2023high}. The purpose of this research direction is twofold. First, and most obviously, research into fault-tolerant compilation and the development of fault-tolerant compilers serves to prepare us to execute quantum algorithms fault-tolerantly on future quantum computers. Secondly, and more immediately, these efforts aid us in estimating resource costs for and evaluating the potential utility of future quantum computers, which is important for informing current directions in quantum computing research.  

With this in mind, we argued in Ref.~\cite{LeBlond2023ASCR} for the importance of developing extensible fault-tolerant compilation platforms with verification. One purpose for doing so is that the resource overheads for different fault-tolerant implementations (e.g. using different instruction sets or even QECC) of quantum algorithms could then be compared, aiding in the optimization of the fault-tolerant layer of the stack. The open-source Lattice Surgery Compiler (LSC) offers a starting point in this direction for surface code lattice surgery specifically~\cite{watkins2023high}. The LSC was originally designed for sequential Pauli-based computation, which is a surface code compilation framework that requires the global re-write of a logical circuit into sequential Pauli product measurements with support over the whole qubit register~\cite{litinski2019game}. However, it also supports direct Clifford+T compilation using techniques from Refs.~\cite{fowler2012surface, horsman2012surface, litinski2018lattice, litinski2019game}. Direct Clifford+T compilation is sometimes considered preferable to sequential Pauli-based computation due to its ability to partially preserve parallelism from the logical circuit in the fault-tolerant layer~\cite{beverland2022surface}. In this work, we have taken advantage of the extensibility of the LSC to directly compile Clifford+T circuits into a lattice surgery instruction set similar to the one proposed in Ref.~\cite{beverland2022surface}, using its Bell-based methodology to compile long-range gates. We have chosen to implement a \textit{local} lattice surgery-based surface code instruction set because of its easy reduction into a limited set of primitive hardware circuits acting independently on one or two hardware tiles that can be readily verified~\cite{leblond2023tiscc}, paving the way toward a verified end-to-end compilation pipeline for quantum circuits. While this choice adds roughly a factor of 2 overhead to some operations, we suspect that this is outweighed by a greater ease in decoding and instruction scheduling in the physical qubit layer compared with the use of extended ancillae to mediate long-range interactions.

The LSC is not only extensible to different instruction sets but also to methods for routing instructions and exploiting their logical parallelism. In the LSC, a \textit{compilation pipeline} refers to the complete process of compiling an input circuit to output lattice surgery instructions, including translations to any intermediate representations. In our upgraded LSC, we have implemented a new compilation pipeline (called the \textit{wave} pipeline) that efficiently detects circuit-level parallelism and combines this information with hardware layout constraints to map logical circuits into time-slices of parallelizable fault-tolerant operations. By effectively handling contention for space by logically parallel operations, the wave pipeline allows the LSC to find a realistic number of logical time-slices (as well as a realistic \textit{active volume}\footnote{The active volume is the active portion of the total space-time volume of the circuit including both the cost of logical operations and the cost of idling logical qubits, since logical errors can accumulate in both. Note that this definition slightly differs from the one introduced in Ref.~\cite{litinski2022active}.}) required to execute a logical circuit fault-tolerantly. Importantly, the wave pipeline implies a change in compilation philosophy as compared with the \textit{streamed} pipeline from the original LSC that was described in Ref.~\cite{watkins2023high}. Namely, the wave pipeline assumes that compilation takes place ahead of circuit execution while the streamed pipeline assumes that compilation takes place in tandem with circuit execution. We will discuss the implications of this choice throughout the paper, but the main benefits of the former approach are that it enables greater exploitation of circuit-level parallelism as well as the \textit{post-hoc} optimization of resources dedicated to magic state distillation and storage, with drawbacks including potential limitations on circuit size due to classical memory constraints. We believe that future work should combine these approaches.

Though many recent efforts towards the resource estimation of practical quantum circuits do consider the costs of fault-tolerance and are often quite sophisticated, they typically stop short of explicit fault-tolerant compilation~\cite{fowler2012surface, litinski2019game, gouzien2021factoring, chamberland2022cat, chamberland2022universal, beverland2022assessing, kim2022fault}. Ref.~\cite{beverland2022assessing} introduced a particularly flexible resource estimator that nevertheless makes several important concessions, such as assuming a constant rate of magic state consumption, using a worst-case formula for synthesis of single-qubit rotations, and neglecting that unused tiles will not contribute to the logical error rate. We believe that explicit compilation within the fault-tolerant layer is necessary to accurately assess the impact of such factors on resource requirements. In this paper, we combine elements of the framework from \cite{beverland2022assessing} with the output of our revised LSC to estimate resources for large, practical quantum circuits. To this end, we explicitly synthesize single-qubit rotations and use the active volume of the final (compiled) circuit to calculate the total logical error rate. We also present a novel method to estimate the resource requirements for magic state distillation and storage. First, the cadence of magic state requests from the compiled output is used to determine the minimum number of magic state factories needed to keep up with magic state consumption in the circuit. Then, knowing the magic state production and consumption rates, the number of magic states in storage after each distillation cycle can be tracked. We calculate contributions to space, time, and active volume from the logical, distillation, and storage portions of the quantum computation and optimize the total of each (or any) of these resources over the code distances under the constraint that the total error rate lies beneath a threshold. As a demonstration of our methodology, we present resource estimates for one Trotter step of a novel ground-state estimation circuit~\cite{kornellsome}. We find that using a minimal number of continuously operating magic state factories that meets the consumption requirements of the circuit causes the spatial cost of magic state storage to dominate, which in turn causes huge total space-time volumes. To bring storage costs down, we propose a method to match the real-time magic state production rate to the consumption rate by turning factories off and on as needed. Using this technique, we find more than three orders of magnitude improvement in space-time volume for our largest circuit. Finally, though we consider an abstract model of quantum computing hardware in this paper, we show that the output of our resource analysis (expressed in terms of resources from the fault-tolerant layer such as hardware tiles and logical time-slices) can be readily combined with hardware-specific resource analyses to produce resource costs in terms of space and time\footnote{The resource estimation pipeline considered here was designed to be compatible with the trapped-ion surface code compiler (TISCC)~\cite{leblond2023tiscc}.}, which completes an end-to-end resource estimation tool-chain.

This paper is divided as follows. In Section~\ref{sec:lattice_surgery}, we discuss lattice surgery compilation and the improvements that we have made to the LSC. In Section~\ref{sec:performance}, we present details on the performance of the LSC. In Section~\ref{sec:resource_analysis}, we discuss our approach to resource analysis using output from the LSC. In Section~\ref{sec:final_estimates}, we discuss resource estimates for a ground state estimation circuit. Finally, we conclude with a summary and outlook in Sec.~\ref{sec:conclusion}.

\subsection{Contributions}

\begin{itemize}
    \item We present the first end-to-end resource estimation pipeline for quantum circuits that utilizes explicit lattice surgery compilation to obtain (a) a realistic number of logical slices, (b) a realistic active volume, and (c) a setting for the optimization of magic state distillation and storage based upon ``real-time'' magic state consumption rates.
    \item To accomplish this, we have made advances to the state-of-the-art in lattice surgery compilation tooling:
    \begin{itemize}
   \item We improve the open-source Lattice Surgery Compiler (LSC)~\cite{watkins2023high} to detect logical parallelism and translate it into parallelism of fault-tolerant operations while handling contention for routes and resource states.

    \item We make other improvements to the LSC such as improved layout generation, more flexible magic state handling, and an improved output format that facilitates its integration within our tool-chain.
    \item We present a detailed performance analysis using both randomly-generated and practical
    circuits. We find that the scalings of both compilation and computation time are consistent with theoretical predictions and compare the performance characteristics of layouts with different ratios of data and ancilla qubits.
    \end{itemize}

     \item Our pipeline paves the way toward verified end-to-end (logical-to-hardware) compilation for fault-tolerant processors through implementing a \textit{local} lattice surgery compilation layer that connects logical lattice instructions to the finite set of logically verified lattice surgery primitives acting on pairs of hardware tiles from Ref.~\cite{leblond2023tiscc}.
    \item We demonstrate the usage of our pipeline by generating resource estimates for the execution of ground state estimation circuits on a hypothetical trapped-ion processor design. We also explore the distillation-storage trade space for these circuits and mitigate the large magic state storage requirements that we discover.
\end{itemize}

\section{Lattice Surgery Compilation}
\label{sec:lattice_surgery}

\begin{figure}[htbp]
    \centering
    \includegraphics[width=0.9\textwidth]{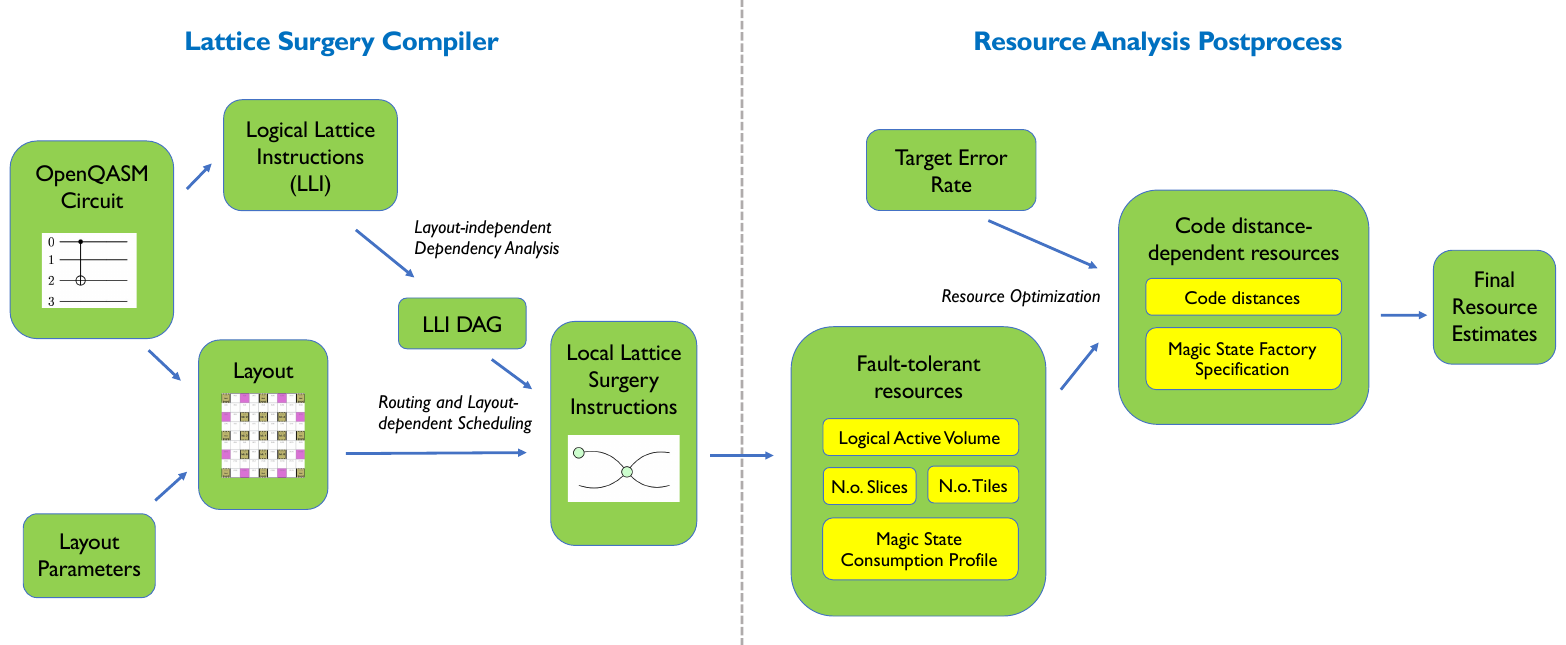}
    \captionsetup{skip=10pt}
    \caption{High-level overview of the Lattice Surgery Compiler (LSC) and its integration with our resource estimation methodology.}
    \label{fig:pipeline}
\end{figure}

As mentioned in Sec.~\ref{sec:introduction}, the primary goal of this work is to enable accurate estimation of the resources required to run large quantum circuits on an abstract model of quantum computing hardware. For a high-level overview of our methodology, see Fig.~\ref{fig:pipeline}. The left-hand side of the diagram, which pertains to explicit lattice surgery compilation within the logical block\footnote{The logical block is the portion of the quantum computer dedicated to executing the logical circuit and excludes resources dedicated to magic state distillation and storage.}, yields total estimates of resources from the fault-tolerant layer such as logical time-slices, hardware tiles, and active volume\footnote{For lack of a better term, we will call these abstract resources \textit{fault-tolerant} resources.} that figure into the resource analysis scheme represented by the right-hand side of the diagram. The purpose of Sec.~\ref{sec:lattice_surgery} is to provide details for our approach to lattice surgery compilation, while the costs of generating and managing magic states, as well as the method for integrating these with logical execution costs, will be reserved for Sec.~\ref{sec:resource_analysis}. First, in Sec.~\ref{sec:lattice_surgery_general}, we will share some general considerations about lattice surgery compilation. In Sec.~\ref{sec:compilation_strategy} we outline our strategy for lattice surgery compilation, and in Sec.~\ref{sec:LSC_revisions} we discuss the LSC, the open-source software package that we have built upon, as well as some specific ways in which we have modified the LSC to suit our compilation strategy. 

\subsection{General Considerations}
\label{sec:lattice_surgery_general}

In this section, we provide background about lattice surgery compilation strategies in order to emphasize the multitude of available choices and frame our set of choices that will be discussed in Sec.~\ref{sec:compilation_strategy}. While we do target one set of choices in this work, our philosophy is to extend the capabilities of an existing open-source compiler rather than build our own. Since there does not seem to be any consensus in the literature as to the best compilation strategy, we hope that this decision paves the way toward future work where resource estimates from different compilation strategies can be effectively compared so that better consensus can be obtained. 

Multiple layers of complexity become apparent while considering fault-tolerant compilation~\cite{LeBlond2023ASCR}. Layers of the stack can be coarsely divided into the logical layer, in which a quantum circuit is expressed by logical gates acting on logical qubits; the fault-tolerant layer, in which logical gates have been re-expressed in terms of fault-tolerant operations on a QECC; and the hardware layer, in which fault-tolerant operations are expressed in terms of native hardware operations. However, these distinctions are not concrete.  Not only can these layers of the stack be broken down further, choices within one layer may be co-dependent with choices made in other layers. To take one example, the specifics of how to implement a logical instruction within the fault-tolerant layer depend on the arrangement of logical resources on the underlying hardware architecture.  This layout can be treated as the subject of optimization by a fault-tolerant compiler, but may also be influenced by hardware limitations such as the need to divide resources across multiple processors. 


For surface codes, there has already been much discussion in the literature about compilation strategies; Refs.~\cite{horsman2012surface,herr2017lattice, herr2017optimization, brown2017poking,fowler2018low,litinski2018lattice,litinski2019game,beverland2022surface,beverland2022assessing, chamberland2022universal,chamberland2022circuit} represent a sub-set of the literature pertaining to surface code compilation using lattice surgery specifically. We will not comprehensively review this literature here, but will briefly discuss a few nuances which serve to illustrate our points above. To provide a setting in which to perform lattice surgery compilation, the arrangement of logical resources on the hardware is often specified in terms of an abstract layout of tiles on a grid (see Fig.~\ref{fig:EDPC_Layout} for examples). These tiles are an abstraction of the area of hardware required to support a single surface code patch at the minimum code distances required by the computation\footnote{See Ref.~\cite{leblond2023tiscc} for the more precise definition of logical tile that is compatible with this work.}. Given a layout specification, the implementation of a logical instruction within the fault-tolerant layer can depend on the availability and locations of surface code patches encoding resource states and/or the availability of connected paths of free tiles (routes) between surface code patches encoding logical data. Once a resource state and/or route is found for a given instruction, one or several ancilla patches may need to be prepared to facilitate the interaction\footnote{Surface code compilation techniques, especially those that use lattice surgery, typically rely on the preparation of surface code patches encoding ancillae that were not present in the original logical circuit. The spatial overhead for ancillae can be significant~\cite{litinski2019game, beverland2022surface, chamberland2022universal}.}. Therefore, the compilation of logical gates into fault-tolerant operations on the surface code is often broken into two stages. In the first stage, each logical gate is compiled into elements of an intermediate representation containing instructions that pertain to the fault-tolerant layer but are still abstracted from layout specifics. In the second stage, the availabilities of resource states and routes are accounted for to implement these abstract instructions in terms of fault-tolerant operations acting on specific hardware tiles~\cite{watkins2023high}. While it has been shown that layout optimization in lattice surgery compilation can be NP hard~\cite{herr2017lattice, herr2017optimization}, authors frequently hand-optimize layouts to suit particular compilation schemes so that resource overheads resulting from routing instructions in the second compilation stage are minimized\footnote{See Refs.~\cite{litinski2019game} and~\cite{litinski2019magic} for vivid examples of hardware layouts being hand-designed to suit a lattice surgery compilation scheme while exploring various space-time trade-offs.}.

In some strategies, such as those presented in Refs.~\cite{litinski2019game, chamberland2022universal,beverland2022assessing}, instead of compiling logical circuits into fault-tolerant operations gate-by-gate, a global re-write of the original logical circuit first eliminates Clifford gates, resulting in a circuit that can be reduced to Pauli product measurements with support over the whole qubit register (these operations are native to surface code lattice surgery~\cite{fowler2018low}). The details of this circuit re-write can depend on one's choice of allowed Pauli product measurements as well as the hardware layout~\cite{litinski2019game, beverland2022assessing, chamberland2022universal}. 
Furthermore, the set of allowed Pauli product measurements can depend on the relative ease with which one can implement twist defects on their underlying hardware, which influences whether the logical layer is re-written to include Pauli product measurements with Y Paulis~\cite{chamberland2022circuit, chamberland2022universal}. Therefore, the hardware layer of the compilation stack has implications that propagate all the way up to the logical layer. 
As mentioned in Sec.~\ref{sec:introduction}, direct Clifford+T compilation is sometimes preferred to the Clifford elimination strategy (sometimes called sequential Pauli-based computation) due to its ability to preserve logical circuit-level parallelism. Within this context, there are also many choices e.g. whether to implement a CNOT gate using extended ancilla patches as in Ref.~\cite{litinski2018lattice} or using long-range Bell states as in Ref.~\cite{beverland2022surface}, whether to implement an S gate using teleportation as in Ref.~\cite{beverland2022surface} or by braiding twist defects as in Ref.~\cite{brown2017poking}, or in how to perform a patch rotation following a transversal Hadamard gate (see Refs.~\cite{vuillot2019code, litinski2019game, beverland2022surface, horsman2012surface, fowler2018low} for different protocols); the optimal choice will be similarly interdependent with choices of layout and hardware constraints. 

These considerations have motivated the realization that the three mentioned layers of the stack are highly inter-dependent and their co-optimization is formidable. At the same time, because the fault-tolerant layer is expected to dramatically increase the resource overhead of quantum computing, any end-to-end resource estimation pipeline for quantum circuits ought to include reasonable choices for it. To summarize this section, we take note that choices must be made in (at least) the following places: (a) a logical gate set with which to express input circuits (e.g. Clifford+T or Pauli product rotations), (b) a lattice surgery-native logical instruction set to translate input circuits to (e.g. the allowed set of Pauli product measurements), (c) an abstract layout specifying regions on the hardware dedicated to storing logical qubits, routing long-range instructions, and generating resource states (d) a protocol for implementing the lattice surgery-native logical instruction set using the layout (including methods for scheduling and routing these instructions as well as generating and managing any required resource states), and (e) a protocol for further compiling routed and scheduled lattice surgery instructions into native hardware gates. As we are convinced that extensible compilation and resource estimation platforms would greatly aid the community in optimizing resources within the fault-tolerant layer and culling the spectrum of propositions for surface code compilation strategies, we present one set of choices in this paper and expand the functionality of the LSC to suit.

\subsection{Overview of Compilation Strategy}
\label{sec:compilation_strategy}

As mentioned in Sec.~\ref{sec:introduction}, we choose Clifford+T for our input logical instruction set and do not perform any global circuit re-write. Within the scope of this paper, it is assumed that the compilation of a quantum algorithm into a logical circuit has already taken place. For an output surface code instruction set, we have chosen a local lattice surgery instruction set based upon the one from Ref.~\cite{beverland2022surface} but with refinements from Ref.~\cite{leblond2023tiscc}. We also make use of an intermediate representation that allows us to translate logical gates into a layout-independent instruction set within the fault-tolerant layer before routing and ordering operations into logical time-slices (slicing) takes place. In this section we will provide an overview of all three of these instruction sets as well as our choice of layout. See Table~\ref{tab:lli} for a summarization of our logical gate set and the associated elements of our intermediate instruction set.

\subsubsection{Circuit-Level Logical Instruction Set}

Here, we will briefly describe the input gate set that we have in mind and overview the approach to compilation for each gate. We allow the Clifford+T gate set; specifically, X, Y, Z, S, $\mathrm{S}^{\dagger}$, T, $\mathrm{T}^{\dagger}$, CNOT, H, and Reset gates can be used as input. The Reset gate nominally clears a logical qubit, and is simply passed directly to output without consuming any space-time volume. Pauli gates (X, Y, and Z) are treated similarly because they can be tracked in software and later combined with measurements according to the Pauli frame tracking methodology~\cite{fowler2012surface, riesebos2017pauli}\footnote{We do not include functionality for tracking the Pauli frame from this information because it has no bearing on resource estimation, which is our ultimate goal for compilation. Besides, the actual run-time Pauli frame will also depend on measurement outcomes gathered during circuit execution.}. We compile both S/$\mathrm{S}^{\dagger}$ and T/$\mathrm{T}^{\dagger}$ gates using standard gate teleportation circuits from Ref.~\cite{fowler2012surface}. The S gate teleportation circuit that we use is ``catalytic'' in the sense that Y states input to the protocol are re-produced by the circuit\footnote{In the ensuing analysis we will neglect the resource cost of Y state generation because we assume it to be negligible.}. Because S and $\mathrm{S}^{\dagger}$ gates utilize the same circuit and the only difference relies on a Pauli Z correction (which is not explicitly applied), we treat them the same within our compiler. Similarly, we treat T/$\mathrm{T}^{\dagger}$ the same because they utilize the same teleportation circuit except for the application of a corrective S gate, which we always explicitly compile. In opposition to the streaming philosophy underlying the original LSC~\cite{watkins2023high}, we conceive of the whole circuit being compiled in advance of execution in order to aid in our efforts toward parallelism and resource optimization. Thus, because in the cases of both T and $\mathrm{T}^{\dagger}$ the corrective gate is required 50\% of the time, we always allocate the required active volume in our compilation and imagine that the system executing the compiled program uses run-time measurement outcomes to apply corrective operations or not\footnote{While this may seem like a downside, this choice significantly improves our ability to detect and compile parallel operations (see Sec.~\ref{sec:LSC_revisions}) and optimize resources dedicated to magic state distillation and storage (see Sec.~\ref{sec:resource_analysis}).}. As mentioned, CNOT gates, which these gate teleportation circuits also rely on, are performed using the Bell-based protocol of Ref.~\cite{beverland2022surface} (to be discussed). Lastly,  H gates are always followed by a patch rotation in order to preserve the orientation of patches' logical X and Z operators throughout the computation, which allows us to comply with the definition of logical tile offered in Ref.~\cite{leblond2023tiscc}.

\subsubsection{Gate-by-gate Processing into Intermediate Representation}

\label{sec:local_instructions}
\begin{table*}[t]
  \centering
  \caption{Circuit-level logical instruction set used in this work together with the logical lattice instructions (LLI) involved in the compilation of each instruction. Where possible, we show the numbers of tiles and slices that each gate is ultimately compiled to. Wherever BellBasedCNOTs are involved, the number of tiles is route-dependent. The details given for  T/$\mathrm{T}^{\dagger}$ gates do not include those of the required corrective S gate.}
  \begin{tabular}{|l|p{8cm}|c|}
    \hline
    \textbf{Logical Gate} & \textbf{Involved Logical Lattice Instructions (LLI)} & \textbf{N.o. Tiles, N.o. Slices} \\
    \hline
Reset & Reset &  0, 0 \\
X, Y, Z & XGate, YGate, ZGate & 0, 0 \\
H & HGate, RotateSingleCellPatch & 2, 3 \\
CNOT & BellBasedCNOT & Route-dependent, 2 \\
T/$\mathrm{T}^{\dagger}$ & MagicStateRequest, BellBasedCNOT, SinglePatchMeasurement & Route-dependent, 2 \\
S/$\mathrm{S}^{\dagger}$ & YStateRequest, BellBasedCNOT, HGate, RotateSingleCellPatch & Route-dependent, 10 \\
    \hline
  \end{tabular}
  \label{tab:lli}
\end{table*}

The first stage of lattice surgery compilation maps the Clifford+T gate set into an abstract instruction set which acts as an intermediate representation within the fault-tolerant layer. This instruction set is not to be confused with the local lattice surgery instruction set that will compose the final compiled output; the key difference between them is that the intermediate representation acts on logical qubit (patch) IDs while the local instructions act on tiles. Being an extension of what were called logical lattice instructions (LLI) in Ref.~\cite{watkins2023high}, we will continue use the term LLI to refer to this intermediate representation for the sake of continuity. Due to the nature of lattice surgery compilation requiring the preparation of ancillae and usage of resource states that were not present in the original logical circuit, new patch IDs need to be generated throughout this stage of compilation. These patch IDs remain virtual until bound with patches instantiated on tiles in the second stage of compilation. It is worth noting here that only a logical dependency graph of LLI results from the first stage of compilation, while the product of the second stage is LLI organized onto logical time-slices (\textit{sliced} LLI) with their compilation into local instructions given in brackets (if applicable). The former depends on dependencies of LLI alone while the latter depends upon the availability of routes, resource states, and the busy status of patches.

We will now overview how each of the logical gates mentioned in the previous section map into LLI (or sequences of LLI) in the first stage of compilation (summarized in Table~\ref{tab:lli}). Reset and Pauli X, Y, and Z are themselves treated as LLI. CNOT gates are simply compiled into BellBasedCNOT LLI, which act on the control and target patch IDs. We note that, while local compilation of the BellBasedCNOT LLI requires the preparation and subsequent measurement of many new logical qubit patches, we do not assign patch IDs to them because they do not have any impact on the logical dependency graph and thus their assignment is unnecessary. H gates are themselves treated as LLI but are followed by a RotateSingleCellPatch LLI. S/$\mathrm{S}^{\dagger}$ gates begin with a YStateRequest LLI acting on a newly generated patch ID (later to be associated with the nearest patch containing a Y state). Then, the appropriate sequence of LLI to execute the required H and CNOT gates according to the catalytic S gate circuit in Ref.~\cite{fowler2012surface} are added. Lastly, T/$\mathrm{T}^{\dagger}$ gates begin with a MagicStateRequest LLI acting on a newly generated patch ID (later to be associated with the nearest patch containing a magic state) followed by the BellBasedCNOT LLI and SinglePatchMeasurement LLI required to implement the gate teleportation circuit from Ref.~\cite{fowler2012surface}. The LLI required by the corrective S gate then follow.

\subsubsection{Further Processing of Intermediate Representation}

The second stage of lattice surgery compilation involves the routing and slicing of the LLI resulting from the first stage. In order to translate the position-agnostic intermediate representation into instructions acting on hardware tiles, the compiler must allocate an appropriate space-time volume to each instruction while avoiding resource conflicts. The layout, which is configurable by input to the compiler (Fig.~\ref{fig:pipeline}), provides a set of constraints at the outset of compilation. Subsequently processed LLI modify these constraints by instantiating patches on tiles along routes, binding or un-binding resource states with IDs, and/or changing the active status of patches and their edges during lattice surgery operations. Once no further LLI can be compiled onto a slice the compiler advances to the next slice, which can also modify the constraints by e.g. re-setting tiles after measurements or re-supplying magic states according to the compilation settings. Given our input instruction set and intermediate representation, our compiler binds patch IDs to available magic states and Y states, allocates busy tiles for RotateSingleCellPatch LLI, and compiles BellBasedCNOT LLI into local lattice surgery instructions in the second compilation stage. 

\begin{figure}
    \centering
    \begin{minipage}{0.45\linewidth}
        \centering
        \includegraphics[width=0.9\linewidth]{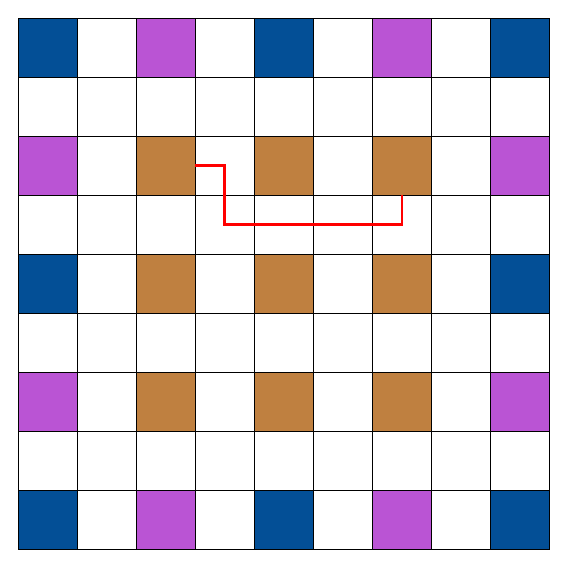}
        \caption{The path corresponding to a BellBasedCNOT LLI\\ (see Fig.~\ref{fig:EDPC_Layout} for layout description).}
        \label{fig:cnot-path}
    \end{minipage}
    \begin{minipage}{0.45\linewidth}
        \centering
        \includegraphics[width=0.9\linewidth]{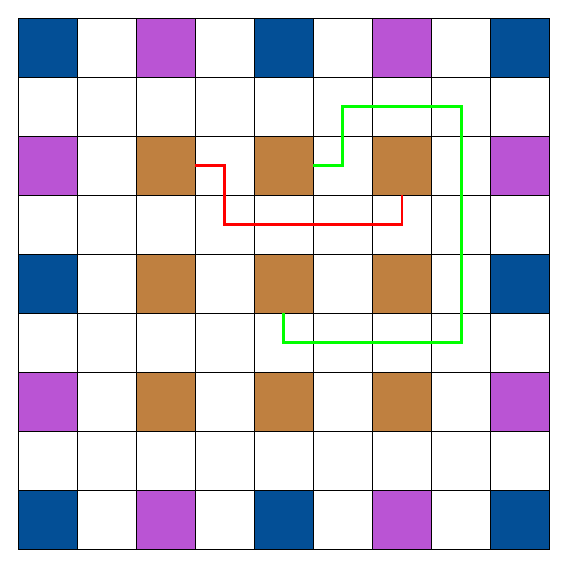}
        \caption{The same as Fig.~\ref{fig:cnot-path} except with a second BellBasedCNOT LLI whose route avoids the first.}
        \label{fig:two-cnot-paths}
    \end{minipage}
\end{figure}

We use results from Ref.~\cite{beverland2022surface} to compile BellBasedCNOT LLI into local instructions (see Table~\ref{tab:local_instructions}). In that protocol, the preparation and usage of a long-range Bell state requires the usage of every tile located along a free route between the control and target patches for two logical time-slices (see Fig.~\ref{fig:cnot-path} for an example of a BellBasedCNOT route), which involves two layers of lattice surgery operations to produce local Bell states and perform Bell measurements between their end-points\footnote{The specific local instructions that the BellBasedCNOT are compiled to depend on whether there is an even or odd number of tiles along the route between the control and target patches, according to Fig. 19a and c in Ref.~\cite{beverland2022surface}.}. While this method involves more local lattice surgery operations than do CNOT compilation protocols involving extended ancilla patches~\cite{horsman2012surface, litinski2018lattice}, it does not have increased resource overhead relative to them. Implementation details for this protocol will be given in Sec.~\ref{sec:LSC_revisions}. We re-iterate that our intention in compiling to a local instruction set rather than using extended ancilla patches is that it facilitates a simple reduction into primitives that allows for easier end-to-end compilation and circuit verification~\cite{leblond2023tiscc}.

\subsubsection{Contention for Space}

Operations that are logically parallel may not be parallelizable within the fault-tolerant layer due to geometrical constraints. For example, a CNOT gate between two qubits requires an ancilla path between the control and target, a T gate requires an ancilla path from a logical qubit to a magic state, and a Hadamard gate requires an uninitialized neighboring ancilla tile so that a patch rotation may be performed. See Fig.~\ref{fig:two-cnot-paths} for a simple example of resource contention between two CNOT gates. Ref.~\cite{beverland2022surface} has proposed one method for retaining logical parallelism in the fault-tolerant layer by finding maximal edge disjoint path sets. Among other things, they found that a layer of $k$ parallel CNOT gates needs depth $\Omega(\sqrt{k})$ to execute on a surface code architecture in the worst-case. In our compilation strategy, we exploit logical parallelism within the fault-tolerant layer by relying on a simple heuristic algorithm (A*) to find vertex disjoint paths and greedily allocating these paths to slices (see Algorithm~\ref{alg:wave} in Sec.~\ref{sec:LSC_revisions} for details). In Sec.~\ref{sec:performance} we will show that, in our implementation, the number of slices scales similarly to these expectations from Ref.~\cite{beverland2022surface}.


\subsubsection{Choice of Surface Code Layout}
Within a lattice surgery paradigm, surface code patches can be simplistically thought of as rectangular entities that are merged and split through edge-wise interactions to generate entanglement or transport quantum information. This results in a type of quantum circuit layout problem, where in general the layout of qubits on a constrained hardware architecture is optimized to reduce circuit depth~\cite{paler2023machine}. In surface code compilation, layouts of equally-sized tiles on a two-dimensional grid are often assumed~\cite{litinski2019game, beverland2022surface}, though it is not always beneficial to have only one tile size available throughout a computation~\cite{litinski2019magic, beverland2022assessing}. Equal X and Z code distances are also not necessarily beneficial in the realistic setting of biased noise~\cite{chamberland2022universal}. In proposals for quantum computer architectures, there is a trade-off between compression of layouts in space and the accessibility of surface code patch edges, as compression in space can lead to overhead in time due to the need to rotate patches to expose needed operators~\cite{litinski2019game}. While keeping the number of tiles dedicated to routing space minimal under the condition that all patch edges are exposed is generally considered to be desirable in sequential Pauli-based computation~\cite{litinski2019game, beverland2022assessing, chamberland2022universal}, it is not as clear in the case of direct Clifford+T compilation where liberal routing space can aid in parallel compilation of logically parallel gates~\cite{beverland2022surface}. In this paper, we take the definition of logical tile presented in Ref.~\cite{leblond2023tiscc} and make use of a family of layouts related to the one from Ref.~\cite{beverland2022surface}\footnote{We will call this family of layouts \textit{EDPC layouts} because of the paper that proposed them.}. We assume that tile sizes are uniform within lattice surgery compilation since, while we do use two levels of distillation with two different code distances, we do not explicitly compile distillation circuits (as explained in Sec.~\ref{sec:resource_analysis}). 

\subsection{Our Revisions to the Lattice Surgery Compiler}
\label{sec:LSC_revisions}
\label{sec:LSC_Review}

Much of the functionality needed to implement our compilation strategy from Sec.~\ref{sec:compilation_strategy} did not originally exist in LSC. However, we leveraged the extensibility of the LSC framework to improve layout generation including the handling of magic states, expand the options for processing logical gates into the intermediate representation (LLI), create a compilation pipeline to handle DAG-based parallelism in the LLI layer, and implement a new compilation layer for processing LLI into local instructions. In this section, we will provide details about our revisions to the LSC in a few areas where we feel that more specificity is warranted than what was presented in Sec.~\ref{sec:compilation_strategy}. We will first review some of the functionality of the original LSC from Ref.~\cite{watkins2023high} in order to provide context for our contributions to it. 

\subsubsection{Lattice Surgery Compiler Review}

The LSC is an open-source software package that compiles logical circuits into lattice surgery instructions on surface code patches. Where we refer to the LSC, we actually refer to the high-performance C++ slicer rather than the smaller-scale Python slicer with simulation\footnote{This package, also known as liblsqecc, can be found at https://github.com/latticesurgery-com/liblsqecc.}. In the original version of LSC,
\begin{itemize}
    \item The input is a circuit decomposed into a standard gate-set consisting of the X, Y, Z, CNOT, S, $\mathrm{S}^{\dagger}$, H, T and  $\mathrm{T}^{\dagger}$ gates. These data are encoded in a minimal subset of OpenQASM 2.0. 
    \item The user specifies the tile layout in a text file where characters specify tiles with different roles such as data qubit (`Q'), ancilla (`A') and routing (`r'). Magic state factories are designated by sets of connected distillation tiles labeled by numbers (`0'-`9').
    \item  One magic state is output from each factory after a (globally) specified number of logical time-slices. Multiple factories can operate in parallel, and they can be configured to output magic states at staggered times. If no magic states are available when compiling a T gate, the compiler will advance time slices until a magic state has been produced. Excess magic states accumulate on the tiles surrounding the factory.
    \item Compilation occurs in two stages, with \textit{logical lattice instructions} (LLI) being the intermediate representation for abstract (layout-independent) lattice surgery instructions. The original LSC contained multiple options for compiling both CNOT gates and S gates into LLI but did not include either the the Bell-based CNOT or the catalytic S gate protocols discussed in Sec.~\ref{sec:compilation_strategy}, and did not include the aforementioned BellBasedCNOT or YStateRequest LLI needed to implement them.
    \item Compilation is implemented as a streamed process, so that compilation could be performed in real-time based on input to classical control software while the quantum computer executes previously compiled instructions. The streaming pipeline (which we retained as an option in our revised version) has the benefit of being able to compile circuits that are too large to fit in memory. It also allows for a limited form of parallelism by compiling streamed instructions onto the same slice until it encounters one that cannot be routed. However, the parallelism is limited by the lack of a logical dependency graph prior to time-slicing instructions, and by multi-stage instructions which automatically request new time-slices for follow-up instructions.
    \item The output is either a QASM-like line-by-line list of LLI acting on patch IDs (without slice information) or a large JSON file containing the detailed slice-by-slice output that can be visualized.
\end{itemize}

\subsubsection{Layout Generation}

We have improved the layout generation capabilities within the LSC by implementing three new types of tiles (specified by text characters as described previously). In the new version of the compiler, there are specified tiles that are reserved for magic states (labeled `M') on which magic states are replenished according to the \verb+disttime+ and \verb+nostagger+ parameters, which respectively specify the rate at which magic states are replenished at these tiles and whether or not they are replenished in a staggered fashion. This improvement enables compilation of T/$\mathrm{T}^{\dagger}$ gates to be dissociated from the space-time footprint of magic state distillation, allowing much greater flexibility in resource estimation and optimization. The \textit{post-hoc} resource analysis discussed in Sec.~\ref{sec:resource_analysis} accounts for both the space-time footprint of magic state factories and the cost to maintain distilled state storage banks, both of which are determined after the cadence of sliced MagicStateRequest LLI is known from the LSC output. Although magic state distillation is not specifically compiled in our framework, our analysis of magic state management (generation, storage, and consumption) makes no approximations as to the availability and consumption of these states at any point in the computation. The second new type of tile we have implemented are tiles that store pre-distilled Y states (labeled `Y'). (Previously, they were produced on-the-fly assuming a twist-based protocol~\cite{watkins2023high}, but we have chosen to employ a twist-free instruction set here). Finally, dead tiles (tiles on which no operations can be performed and which are avoided by the router) have also been implemented (labeled `X') so that non-rectangular layouts can be specified (though we have not yet made use of this).

\begin{figure}[t]
\begin{subfigure}{0.33\textwidth}
    \centering
    \includegraphics[width=0.8\linewidth]{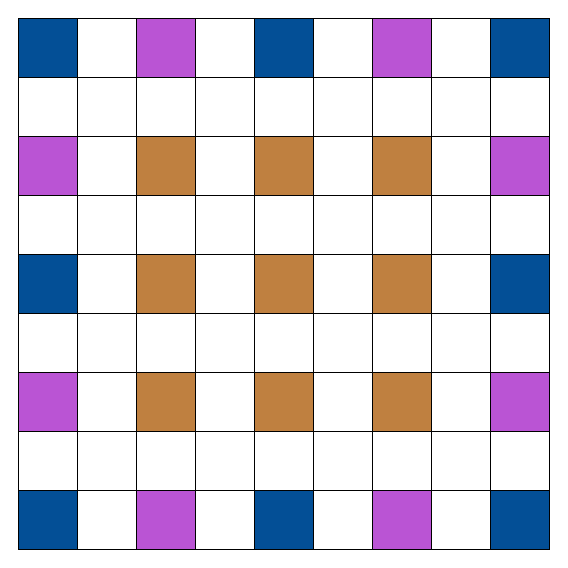}
\end{subfigure}
\begin{subfigure}{0.33\textwidth}
    \centering
    \includegraphics[width=0.8\linewidth]{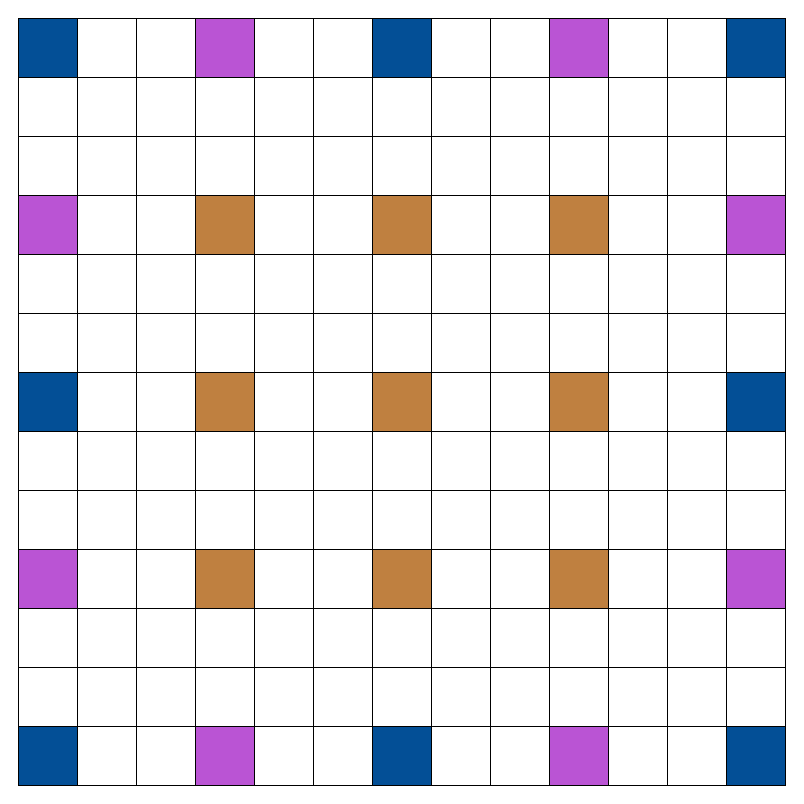}
\end{subfigure}
\begin{subfigure}{0.33\textwidth}
    \centering
    \includegraphics[width=0.8\linewidth]{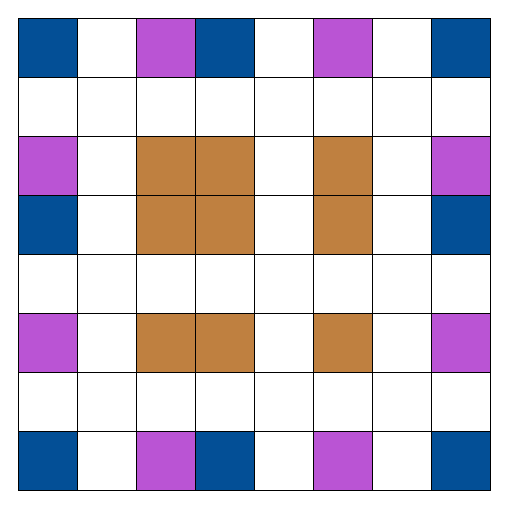}
\end{subfigure}
\captionsetup{skip=10pt}
\caption{EDPC layouts for 9 logical qubits, generated using the (a) default, (b) num\_lanes = 2, and (c) condensed settings. Brown tiles in the bulk correspond to logical qubits. Blue and pink/purple tiles on the boundary correspond to Y states and tiles reserved for magic states, respectively. White tiles indicate tiles dedicated to routing instructions and the preparation of ancillae.}
\label{fig:EDPC_Layout}
\end{figure}

For our resource estimates, we implemented a layout generator that constructs the EDPC layout as illustrated in Fig.~\ref{fig:EDPC_Layout}, accessible through the \verb+-L edpc+ command line option (Fig.~\ref{fig:EDPC_Layout}(a) represents the EDPC layout that is generated by default using this setting). In Fig.~\ref{fig:EDPC_Layout}, the brown tiles in the bulk (bound to patch IDs) correspond to patches encoding logical qubits from the input circuit. Blue and pink/purple tiles on the boundary correspond to pre-distilled Y states and tiles that are reserved for magic states, respectively. While these have no patch IDs by default, they are bound with IDs upon encountering YStateRequest and MagicStateRequest LLI during the second stage of compilation. White tiles indicate ancilla tiles that can be used for routing (in the case of our compilation strategy, this is where long-range Bell states are prepared and where patch rotations take place). In order to observe the effects of different data:ancilla tile ratios, we implemented \verb+num_lanes+ and \verb+condensed+ options to \verb+-L edpc+. The effects of these options on the generated layout are depicted in Fig.~\ref{fig:EDPC_Layout}(b) and (c), respectively. The former option can be used to open up more lanes for routing CNOT gates at the expense of greater spatial overhead, while the latter option decreases the total routing space by blocking off two edges of each logical qubit. The effects of these options on circuit depth and space-time volume will be explored in Sec.~\ref{sec:performance}.

\subsubsection{Routing and Parallelism using the Wave Pipeline}

Our modified version of LSC detects parallel instructions from the logical circuit and attempts to schedule as many instructions on each slice as possible while avoiding contention. The new \textit{wave} pipeline, made accessible through the command line option \verb+-P wave+, uses a greedy scheduling algorithm within the second stage of compilation to route and slice logically parallel operations in ``waves'' according to both their logical dependency graph and constraints imposed by the layout. The wave pipeline constructs a directed acyclic graph (DAG) for the entire layout-independent computation, and includes an As-Soon-As-Possible scheduler that takes routing conflicts into account. (See Algorithm~\ref{alg:wave} for details.) This scheduling algorithm was chosen for simplicity; future versions of the compiler could potentially use alternative scheduling algorithms at this stage. One downside of this approach is the abandonment of the streaming philosophy underlying the original LSC. While we imagine that a windowed-DAG methodology could be used in order to obtain parallelism while not sacrificing the ability to stream instructions, we reserve this for later work. 

\begin{algorithm} [t]
\caption{Wave pipeline scheduling algorithm}
\label{alg:wave}
\begin{algorithmic}
    \State Construct DAG of instructions and dependencies
	\State $currentwave \gets ( \text{list of instructions with no dependencies} )$
	
	\While {there are still instructions to schedule}
		\State $nextwave \gets \emptyset$
		
		\While {$currentwave \neq \emptyset$}
			\State $instruction \gets \text{head of } currentwave$
			\If {can find necessary resource(s) (e.g. available route and/or magic state) $s$ to slice  $instruction$}
				\State Schedule $instruction$ using $s$
				\For {$i$ in instructions whose dependencies are now met}
					\State Append $i$ to $currentwave$
				\EndFor
			\Else
				\State Transfer $instruction$ to $nextwave$.
			\EndIf
		\EndWhile
	
        \State $currentwave \gets nextwave$
	\EndWhile
	    
\end{algorithmic}
\end{algorithm}

\subsubsection{Local Compilation Layer}
\label{sec:local_instructions}
\begin{table*}[t]
  \centering
  \caption{Local instruction set implemented within the second stage of lattice surgery compilation in the LSC. Italicized instructions are not considered part of our instruction set but are included for reference as sub-circuits used by other members. Table adapted from Ref.~\cite{leblond2023tiscc}.}
  \begin{tabular}{|l|p{7cm}|c|c|}
    \hline
    \textbf{Instruction} & \textbf{Description} & \textbf{Tiles In/Out} & \textbf{Logical Time-Slices} \\
    \hline
    TwoPatchMeasure& Measures the joint XX/ZZ operators of two vertically/horizontally-adjacent initialized tiles & 2 & 1 \\
    BellPrepare & Initializes a Bell state on two adjacent uninitialized tiles & 2 & 1 \\
    BellMeasure & Performs a destructive Bell basis measurement on two adjacent (initialized) tiles and makes uninitialized & 2 & 1 \\
    ExtendSplit & A patch extension followed by a split operation & 2 & 1 \\
    MergeContract& A merge operation followed by a patch contraction & 2 & 1 \\
    Move & A patch extension followed by a patch contraction & 2 & 1 \\
    \textit{Patch Contraction} & Contracts one initialized two-tile patch into a one-tile patch while    preserving the encoded state & 2/1 & 0 \\
    \textit{Patch Extension} & Extends one initialized one-tile patch into a two-tile patch while preserving the encoded state & 1/2 & 1 \\
    \hline
  \end{tabular}
  \label{tab:local_instructions}
\end{table*}

Within the second stage of compilation we have implemented a new local compilation layer that utilizes the instruction set found in Table~\ref{tab:local_instructions}. We currently use this local instruction set to compile the BellBasedCNOT LLI, which accounts for most (but not all) of the physical operations output by the LSC within our framework. This layer could in principle be adapted to the compilation of other LLI. As an example, while the RotateSingleCellPatch LLI is currently compiled by merely designating a two-tile busy region for three time-steps according to the protocol in Ref.~\cite{litinski2019game} (this is unchanged from the original version of LSC), it could be broken further into corner movements and other primitives such as those that were explored in Ref.~\cite{leblond2023tiscc}. The instruction set of Table~\ref{tab:local_instructions} is said to be ``local'' because each member acts on vertically or horizontally adjacent hardware tiles. Input tiles are assumed to adhere to the definition given in Ref.~\cite{leblond2023tiscc}, and where initialized, employ a standard stabilizer configuration that implies a directionality to ZZ and XX measurements. The implementation details of these instructions, which depend on whether the targeted tiles are horizontally or vertically separated, are given in Ref.~\cite{leblond2023tiscc}, where they are also explicitly compiled into primitive circuits native to a trapped-ion processor that are verified using simulated tomography protocols.  

\subsubsection{Outputs of the Lattice Surgery Compiler}
In addition to the outputs available in the original LSC, we have implemented the \verb+--printlli sliced+ option which yields a line-by-line (slice-by-slice) list of comma-separated LLI with their associated patch IDs. In the case of the BellBasedCNOT LLI, the further compilation into local instructions over multiple slices appears in brackets along with the pair of tiles on which each local instruction acts. We have additionally enabled the LSC to compute and output certain useful statistics such as the active volume of the logical computation using the slice visitation methodology described in Ref.~\cite{watkins2023high}. 

\section{Performance of the Lattice Surgery Compiler}
\label{sec:performance}

Now that we have described our compilation methodology and its implementation within the LSC, we demonstrate the compiler's performance on random circuits. To this end, we consider both the average compilation time per gate and the preservation of logical parallelism within the fault-tolerant layer to be key performance metrics. In addition to these metrics, we provide some insight into the space-time trade-offs that exist within the EDPC layout family. To generate random circuits, we specify a number of logical qubits and then add random gates until a cut-off on the circuit depth has been met. Our random circuits contain either CNOT gates or a 50/50 mixture of CNOT gates and T gates. We note that, since our compilation methodology heavily relies on gate teleportation circuits (see Sec.~\ref{sec:compilation_strategy}.1), real circuits boil down to CNOT and Hadamard gates. As such, we focus on CNOT gates to demonstrate trade-offs related to routing and parallelism, and mix in T gates to capture both the impact of the requirement to route to boundary tiles to fulfill MagicStateRequests and YStateRequests and the impact of Hadamard gates (from the corrective S circuit) in blocking paths. We consider scaling in two different regimes: in one, we hold the circuit depth constant and generate circuits with numbers of logical qubits up to 10,000, and in the other, we hold the number of logical qubits constant and vary the circuit depth up to 500. At each grid point we generate ten circuit samples and compile them using the LSC. Each compilation task was performed on a single core of the Compute and Data Environment for Science (CADES) high-performance computing cluster at Oak Ridge National Laboratory. 

\begin{figure}[t]
\begin{subfigure}{0.49\textwidth}
    \centering
    \includegraphics[width=\linewidth]{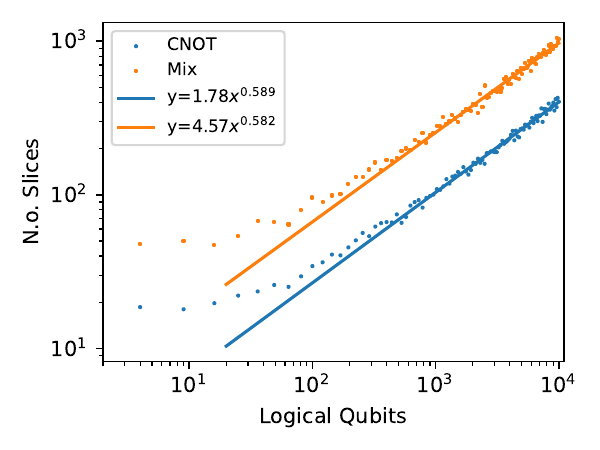}
\end{subfigure}
\begin{subfigure}{0.49\textwidth}
    \centering
    \includegraphics[width=\linewidth]{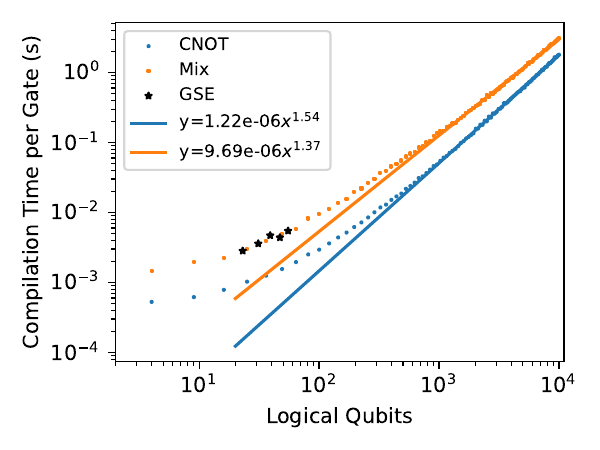}
\end{subfigure}
\captionsetup{skip=10pt}
\caption{Scaling of (a) the number of slices in the compiled output and (b) the average compilation time per gate against the number of logical qubits present in randomly generated circuits of CNOT gates and a 50/50 mixture of CNOT gates and T gates. The circuit depth is kept fixed at 10 and data points are averaged over 10 circuit samples. Trend-lines are power-law fits starting from 1000 logical qubits. In (b), results for the practical GSE circuits from Table~\ref{tab:gse_circuits} are included for comparison.}
\label{fig:LSC_performance}
\end{figure}

In Fig.~\ref{fig:LSC_performance}(a), we plot the average number of slices in the compiled output against the number of logical qubits $x$ with circuit depth fixed to 10. We find that this quantity scales a little faster than $O(\sqrt{x})$, which is consistent with the worst-case expectation from Ref.~\cite{beverland2022surface} (mentioned previously in Sec.~\ref{sec:compilation_strategy})\footnote{While a more direct comparison requires uniform gate densities, we find that the average gate density decays weakly ($\sim x^{-0.1}$) with the number of logical qubits in our random circuits.}. Unsurprisingly, the number of slices increases with the proportion of T gates, which is reflected in the constant factor. In Fig.~\ref{fig:LSC_performance}(b), we plot the compilation time per gate against the number of logical qubits $x$ with circuit depth fixed to 10. The high levels of contention in our randomly generated circuits have a substantial impact on the compile time performance of our implementation. The dominant cost of our implementation comes from the A* routing algorithm, which scales as $O(x \log x)$ in the worst case. This worst case is realized when no paths can be found by A*, and this occurs precisely when there are routing conflicts. To add insult to injury, if a gate cannot be scheduled in slice $t$, the router is re-run for that gate in slice $t + 1$. Consequently, according to the results in Fig.~\ref{fig:LSC_performance}(a), high levels of contention cause the A* router to be called $O(\sqrt x)$ times for each gate. Combining these observations, we expect our compile time per gate to scale as $O( x^{1.5} \log x )$ in high contention scenarios. This prediction is supported by Fig.~\ref{fig:LSC_performance}(b), since we see a little faster than $O(x^{1.5})$ scaling for CNOT circuits up to 10,000 qubits. In the mixed circuits, the slicing of certain instructions may be limited by the availability of resource states rather than by contention for routes, which may explain the lower exponent in this case. We also see that the leading coefficients of our best fit power laws increase with the proportion of T gates due to those gates being compiled into several gates according to our strategy explained in Sec.~\ref{sec:compilation_strategy}. Interestingly, we find that results from the ground state estimation circuits studied in Sec.~\ref{sec:final_estimates} overlap with the mixed circuit results in Fig.~\ref{fig:LSC_performance}(b). 


\begin{figure}[t]
\begin{subfigure}{0.49\textwidth}
    \centering
    \includegraphics[width=\linewidth]{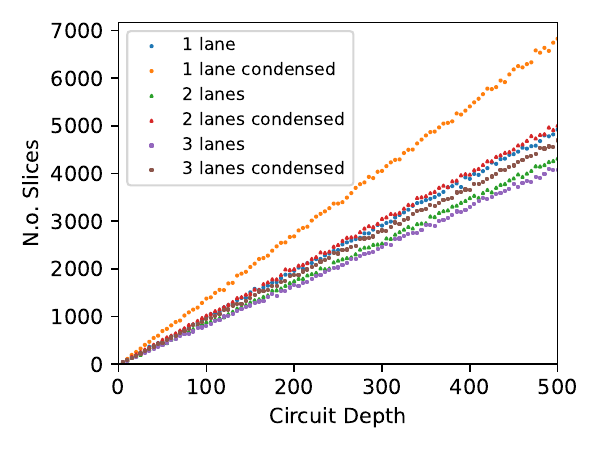}
\end{subfigure}
\begin{subfigure}{0.49\textwidth}
    \centering
    \includegraphics[width=\linewidth]{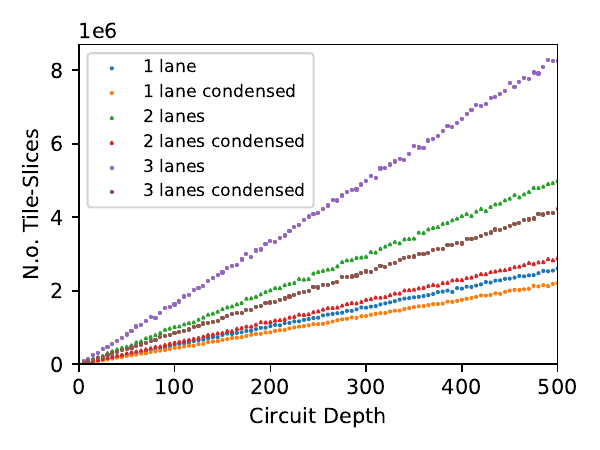}
\end{subfigure}
\captionsetup{skip=10pt}
\caption{The number of slices in the compiled output (a) and the total logical volume (in tile-slices) (b) against the circuit depth in randomly generated circuits of a 50/50 mixture of CNOT and T gates for a variety of layouts within the EDPC layout family. The number of logical qubits is kept fixed at 100 and data points are averaged over 10 circuit samples.}
\label{fig:EDPC_layout_performance}
\end{figure}

\begin{figure}[t]
\begin{subfigure}{0.49\textwidth}
    \centering
    \includegraphics[width=\linewidth]{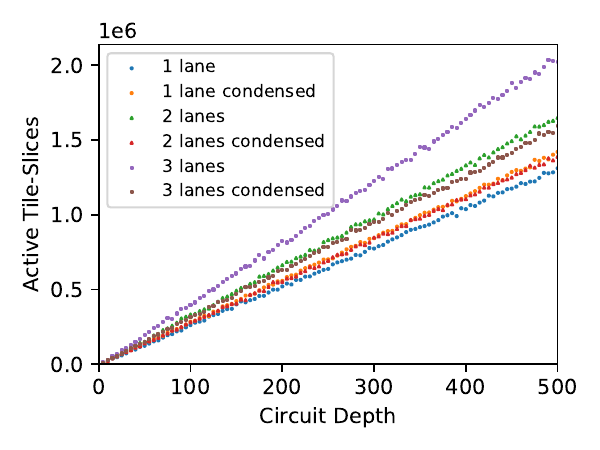}
\end{subfigure}
\begin{subfigure}{0.49\textwidth}
    \centering
    \includegraphics[width=\linewidth]{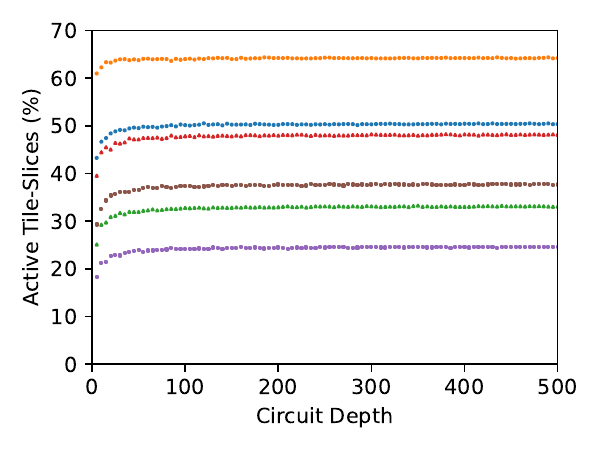}
\end{subfigure}
\captionsetup{skip=10pt}
\caption{The total logical active volume (in tile-slices) (a) and the same expressed as a percentage of the total logical volume (b) against the circuit depth in randomly generated circuits of a 50/50 mixture of CNOT and T gates for a variety of layouts within the EDPC layout family. The number of logical qubits is kept fixed at 100 and data points are averaged over 10 circuit samples.}
\label{fig:EDPC_layout_performance2}
\end{figure}

Next, Figs.~\ref{fig:EDPC_layout_performance} and~\ref{fig:EDPC_layout_performance2} provide insight into space-time trade-offs that exist within the EDPC layout family. We expect the slice performance of these layouts to differ because the layouts generated using different combinations of the \verb+num_lanes+ and \verb+condensed+ parameters have different ratios of data:ancilla tiles. In Fig.~\ref{fig:EDPC_Layout}(a), (b), and (c), for example, we see that there is respectively a 3:1, 8:1, and 5:4 data:ancilla tile ratio in the bulk. Additionally, in condensed layouts, logical qubit tiles have only two free edges (as long as the number of logical qubits is divisible by 4); otherwise there are four free edges. In Fig~\ref{fig:EDPC_layout_performance}(a), it is evident that having more routing lanes leads to a lower number of slices, though the benefit appears to be limited. Also, the condensed option causes a jump in the number of slices in all cases, especially for 1-lane condensed, which yields the highest number of slices by far. In Fig~\ref{fig:EDPC_layout_performance}(b), we examine the effect of layout on the total logical volume of the circuit\footnote{More precisely, we record the total volume (in tile-slices) reported by the Lattice Surgery Compiler while excluding magic state distillation and storage costs; this value is the volume of the logical computation given a constant rate of magic state replenishment at reserved tiles, and is independent of code distance. The real space-time cost of algorithms, including contributions from magic state distillation and storage, will be accounted for in later sections.}. One can see that, in this respect, condensed layouts generally perform better than their un-condensed counter-parts. That said, while 1-lane condensed [Fig.~\ref{fig:EDPC_Layout}(c)] is the best-performing layout by this metric, its performance is similar to that of 1-lane [Fig.~\ref{fig:EDPC_Layout}(a)].  To better distinguish between them, we further evaluate the effect of layout parameters on the logical active volume (see Sec.~\ref{sec:introduction} for the definition of this quantity), which has implications on total logical error and, accordingly, code distance and space-time volume (see Sec.~\ref{sec:resource_analysis}). In Fig.~\ref{fig:EDPC_layout_performance2}, we plot the logical active volume both as an absolute quantity [Fig.~\ref{fig:EDPC_layout_performance2}(a)] and as a percentage of the total volume [Fig.~\ref{fig:EDPC_layout_performance2}b]. We find that, while 1-lane condensed makes most efficient use of its tile-slices by having the highest active volume percentage, the total active volume for 1-lane is lower, yet still similar to that of 1-lane condensed. Ultimately, we conclude that 1-lane and 1-lane condensed are comparably good layouts, and that the choice between them will depend on whether one would prefer to save time or space in their computation. 
We also conclude that the benefit of opening up additional lanes for routing is probably not worth the space cost. Given these conclusions, we have opted to use the 1-lane EDPC layout in our resource analyses in Sec.~\ref{sec:final_estimates}.

\subsection{Outlook for the Lattice Surgery Compiler}
\label{sec:outlook}
In practice, circuits with dozens of qubits and millions of gates can be compiled in hours (see Table~\ref{tab:gse_circuits} in Sec.~\ref{sec:final_estimates}). Thus, larger circuits, especially those involving many qubits, may require decomposition into smaller sub-circuits that are compiled separately. While in this article we focus on whole-circuit compilation, we have found that the performance of the LSC is more than enough to compile key sub-circuits of the ground state estimation circuits that we consider in Sec.~\ref{sec:final_estimates} according to the methodology from Ref.~\cite{kornellsome}. Future work should allow LSC to link these sub-circuits together; we expect that this strategy would allow much larger circuits to be compiled without issue.  

\section{Resource Estimation Methodology}
\label{sec:resource_analysis}
In this section, we discuss how the Lattice Surgery Compiler can be used to generate fault-tolerant resource estimates for quantum circuits.

\subsection{Usage of the Lattice Surgery Compiler for Resource Estimation}

To review, the revised LSC compiles Clifford+T circuits gate-by-gate into a local, tile-based lattice surgery instruction set. The final output is \textit{sliced}, meaning that the output is not merely a list of sequential instructions but is organized into logical time-slices on which instructions are performed in parallel. In this section, we describe how the sliced LSC output (together with its output statistics) can be used to realistically estimate resources and quantify the total logical error rate for quantum circuits. Then, we show how these outputs can be used for resource optimization subject to a total error rate threshold\footnote{This process may also be used as a subroutine that estimates resources of sub-circuits toward the goal of estimating resources of a larger circuit. The methodology for breaking the larger circuit down into sub-circuits, quantifying the resources of each sub-circuit, and recombining results into a total resource analysis is out of scope for this paper.}. Using the sliced output of the LSC, it is easy to extract quantities such as the total number of a particular instruction per slice or the running total of that instruction type. From the output statistics, the useful quantity for the sake of resource estimation is the \textit{active volume}, which will be used to calculate the logical error rate.

As seen in Fig.~\ref{fig:EDPC_Layout}, the layout specification generated by the LSC using the \verb+-L edpc+ option contains tiles that are reserved for magic states which are replenished at a rate given by the \verb+disttime+ option. As alluded to in Sec.~\ref{sec:lattice_surgery}, our resource estimation methodology does not assume a 1:1 correspondence between magic state factories and the tiles reserved for magic states in the layout. Instead, we calculate the necessary number of magic state factories using the rate at which the \verb+RequestMagicState+ LLI is found in the sliced output. The distillation cycle time also depends on the specifics of the distillation factory used, which is also calculated downstream of the LSC output and does not depend on the \verb+disttime+ option. We note that parallelism within the fault-tolerant layer (and therefore algorithm run-time) can improve with a higher rate of replenishing magic states, but in turn increases the number of factories needed to keep up with supply (and therefore the spatial footprint of distillation). Conversely, a longer algorithm run-time or a resulting surplus of reserved magic states can increase the minimum required code distance to remain below a logical error rate threshold, increasing the spatial footprint as well. Due to these non-trivial space-time trade-offs, we expect that \verb+disttime+ is a parameter worthy of optimization. In this work, we make the simplifying assumption that magic states are available as quickly as possible given constraints set by the layout by keeping \verb+--disttime 1+\footnote{Preliminary tests have demonstrated that this parameter does not have a major impact on the final number of slices as long as it remains below a certain cutoff related to the maximum rate at which magic states can be consumed.}. 

\subsection{Cost of Magic State Distillation and Storage}
We primarily follow the magic state factory design framework proposed in Ref.~\cite{beverland2022assessing}. In this framework, the total factory $\mathcal{D}$ is composed of $R$ distillation rounds $r \in \{1, 2, \dots, R\}$, each using $C_r$ copies of a distillation unit $M_r$ operating at code distance $d_r$. We assume a negligible cost of state injection and of transferring the output magic states from one level of distillation to the correct positions and code distances to use in the next level. Though each of the mentioned parameters can be optimized, we make multiple \textit{a priori} decisions regarding them. We assume $R = 2$ levels of distillation where $C_2 = 1$, and we assume that $M$ is a 15:1 distillation unit that has been compiled to the same local instruction set that we use in the LSC (see Fig. 9 and Table VI of Ref~\cite{beverland2022assessing} for details). We constrain $d_2$ to be same code distance as that used in the logical part of the computation (i.e. the \textit{full} code distance). For a given choice of $d_1$, we require that $C_1$ be the smallest number that yields $P_1^{acc}(\geq 15) \times P_2^{acc} \geq P_{th}$, where $P_1^{acc}(\geq 15)$ is the probability of at least 15 output states being accepted at the first level, $P_2^{acc}$ is the acceptance probability at the second level, and $P_{th}$ is a pre-decided threshold (taken to be 0.985 here). The final run-time of the distillation factory is $\tau(\mathcal{D}) \equiv \ceil{\tau_1 + \tau_2}$, where $\tau_r$ is the number of full-distance slices required for distillation round $r$\footnote{Where $d_1 \neq d_2$, $\tau_1 = \frac{d_1}{d_2} \tau_2$ may not be a whole number.}. The total spatial cost of the distillation factory is $n(\mathcal{D}) \equiv \ceil{\max_{r} \left[C_r n_r\right]} = \ceil{\max{\left(C_1 n_1, n_2\right)}}$, where $n_r$ is the number of full-distance tiles required for $M$ at distillation round $r$\footnote{Similarly, we calculate $n_1$ as $n_1 = \frac{2d_1^2 - 1}{2d_2^2 - 1} n_2$}. The active volume of the distillation factory is approximated to be $V(\mathcal{D}) \simeq C_1 n_1 \tau_1 + n_2 \tau_2$. In our resource analysis script, we treat the factory selection process as a function that receives $d_1$ and $d_2$ as parameters and outputs $n(\mathcal{D})$, $\tau(\mathcal{D})$, $V(\mathcal{D})$, and the distilled state error rate $P_T(\mathcal{D})$. This function is used as a subroutine within the resource optimization process to be described.

The number of magic state factories needed to keep up with the rate of magic state requests is calculated using the sliced output from the LSC together with $\tau(\mathcal{D})$. First, consider (positive) integer multiples $k$ of $\tau(\mathcal{D})$ to be the number of slices required to complete distillation cycle $k$. Then, let $m_i$ be the number of magic states consumed in slice $i$ of the compiled output (this is the number of \verb+RequestMagicState+ statements on line $i$ of the sliced output). We calculate the number of consumed magic states through slice $k\tau(\mathcal{D})$ as:  \beq m(k) \equiv \sum_{i=1}^{k\tau(\mathcal{D})} m_i \eeq 
The object is to calculate $N$, which is the number of magic states that need to be produced in each distillation cycle to keep up with the consumption rate. We note that, since our distillation factories output one state each distillation cycle, $N$ is equal to the number of factories. To cover the requests made during the first cycle, we require $w = \ceil{m(1)/N}$ warm-up distillation cycles. Then, we minimize $N$ with respect to the condition $m(k) \leq N(w + k -1)$ because the number of consumed states after $k$ distillation cycles has to be less than or equal to the number produced as of the $(w+k-1)$th cycle at every $k$. Using the expression for $w$, we relax this condition to $m(k) \leq m(1) + N(k-1)$ and solve for N: \beq N \geq \frac{m(k) - m(1)}{k-1} \eeq Since we want to supply the minimum number of states per cycle that satisfies this inequality, we write: \beq N = \ceil*{\max_k \frac{m(k) - m(1)}{k-1}}. \eeq We note that one might also want to include additional warm-up cycles $w_{add}$ to decrease $N$ further. In this case, we replace $k$ with $k + w_{add}$ in the denominator of the above formula (in what follows we call $w_{total} = w + w_{add}$).

It is implied in the above analysis that we have a storage bank of magic states from which the reserved tiles in the layout can be replenished. While we do not count the costs of moving states from factories into banks or from banks to reserved tiles, we do count the spatial and active volume costs of storage, which we assume to be dominant compared to these other contribution. The number of magic states in reserve at the beginning of cycle $k+1$ is $R(k) \equiv N(w_{total} + k) - m(k)$ as long as the total number of magic states needed by the algorithm $m(k_{max})$ has not yet been produced\footnote{$k_{max}$ represents the total number of distillation cycles-worth of slices within the logical computation (not including warm-ups) and $k_{stop}$ is the cycle at which distillation halts.}. For all cycles subsequent to that cycle ($k>k_{stop}$), magic state production halts and $R(k) = m(k_{max}) - m(k)$. The spatial overhead of magic state storage is equal to the maximum number of magic states ever in reserve i.e. $n_{storage} = \max_k R(k)$. The storage contribution to active volume is approximately: \beq V_{storage} \simeq \left(\sum_{k=1}^{k_{max}-1} R(k) + \sum_{j=1}^{w_{total}} jN \right)\tau(\mathcal{D}) \eeq Essentially, the number of magic states in storage at the beginning of every distillation cycle is multiplied by $\tau(\mathcal{D})$ time-steps and the sum is taken over distillation cycles. The second term in the parenthesis accounts for the $w_{total}$ warm-up cycles. $V_{storage}$ is approximate because we do not account for the slice-by-slice consumption of magic states from the reserve but instead coarse-grain by distillation cycle. Using the parameter $k_{stop}$, we can now calculate the total active volume from magic state distillation: \beq V_{dist} = N V(\mathcal{D}) (w_{total} + k_{stop}).\eeq 

\subsection{Total Resource Costs and Calculation of Error Rates}
With all of the above in mind, we are able to count resource costs. The total spatial costs (in hardware tiles) of running the circuit is \beq n_{total} = Nn(\mathcal{D}) + n_{storage} + \left(2\ceil*{\sqrt{L}}+3\right)^2,\eeq where the third term is the spatial contribution from the EDPC layout\footnote{We have used the default settings for the EDPC layout for resource estimation, i.e. 1 routing lane separating all logical qubit patches.} and $L$ is the size of the register from the input circuit to LSC. The total time cost of running the circuit (in logical time-slices) is \beq \tau_{total} = w_{total}\tau(\mathcal{D}) + \tau_{logical},\eeq where $\tau_{logical}$ is the number of slices output from LSC. Lastly, the total active volume is simply additive as in \beq V_{total} = V_{logical} + V_{dist} + V_{storage}. \eeq Using this information, we may estimate error rates for the logical, distillation, and storage parts of the computation. Given a code distance $d$ and a hardware-dependent physical error rate $p$, we calculate the logical error rate of a surface code patch per logical time-slice as $P(d) = 0.1d(\frac{p}{0.01})^{\frac{d+1}{2}}$, where the pre-factor and the threshold value are taken from Ref.~\cite{fowler2018low}. We consider two different sets of estimates for hardware gate infidelities that can be selected through the \verb+param_type+ option in the resource analysis script (for \verb+--param_type current+, $p_1 = 4\times 10^{-5}$ and $p_2 = p_I = 3\times 10^{-3}$~\cite{quantinuumH1}; for \verb+--param_type projected+, we modify these using recent demonstrations: $p_I = 1\times10^{-4}$~\cite{an2022high} and $p_2 = 6\times10^{-4}$~\cite{clark2021high})\footnote{These numbers are based upon current and projected estimates for the infidelity of single-qubit gates ($p_1$), two-qubit gates ($p_2$), and state initialization ($p_I$) in trapped-ion hardware specifically; one could easily modify the script to include parameters pertaining to other hardware modalities. The resource estimates reported in Sec.~\ref{sec:final_estimates} use the projected parameter set.}. Since syndrome extraction circuits are primarily made of CNOT gates~\cite{fowler2012surface}, we use $p=p_2$ in the formula for $P(d)$. Further, since our analysis requires injected magic states as input to the first level of distillation, we calculate the injected state error according to the formula from Ref.~\cite{lao2022magic} (this is a refinement of the state injection protocol of Ref.~\cite{li2015magic} for the rotated surface code), which is $P_I = \frac{3}{5} p_2 + p_I + \frac{2}{3} p_1$.

The logical contribution to error is then $\epsilon_{logical} = V_{logical} P(d)$, where the logical active volume $V_{logical}$ is taken from the statistics output of LSC. The contribution to total error from magic state storage is also simply $\epsilon_{storage} = V_{storage} P(d)$. The distillation contribution to error is $\epsilon_{dist} = m(k_{max}) P_T(\mathcal{D})$ i.e. it is the total number of magic state requests multiplied by the distilled state error (the distilled state error is calculated as in Ref.~\cite{beverland2022assessing}).  The total error of the computation is taken to be the sum of these contributions: \beq \epsilon = \epsilon_{logical} + \epsilon_{dist} + \epsilon_{storage} \eeq The total error $\epsilon$ is required to stay beneath a total error budget $\epsilon_{thr}$ which may depend on the algorithm and whether the circuit in question is the whole circuit or just a sub-circuit of the whole circuit. In our resource analysis script, this threshold is an input given by the \verb+error_budget+ option, and represents the primary constraint to resource optimization. To use the script, one chooses a resource to optimize using the \verb+minimize_what+ flag with options \verb+{space, time, space-time, active_volume}+. While space, time, or space-time are more traditional resources than active volume to optimize, we suspect that in some modalities the active volume could serve as a proxy for energy consumption. The code distances $d_1$ and $d_2$ are parameters being subject to optimization and can be given bounds through the options \verb+min_d1+, \verb+min_d2+, and \verb+max_d2+. The number of additional warm-up cycles $w_{add}$ can also be made subject to optimization by specifying \verb+max_w+, and in Sec.~\ref{sec:final_estimates} we will show the benefit of including these warm-up cycles.

\subsection{Three Approaches to Resource Optimization}
The above analysis (which we refer to as our \textit{default} approach where $w_{add}$ is constrained to be zero) implies a constant rate of magic state production throughout the computation and an unlimited capacity for magic state storage. As demonstrated in Sec.~\ref{sec:final_estimates}, this can result in the vast majority of the quantum computer being dedicated to magic state storage, which is clearly not optimal. We have seen this possibility manifest where the rate of magic state consumption is higher toward the beginning of the circuit than in its bulk, since in that case the number of factories needed to supply a faster rate in the beginning will over-produce states during later, lower-consumption parts of the circuit. This leads to a huge excess of magic states and an early $k_{stop}$, which causes a high $n_{storage}$ and $\epsilon_{storage}$ and greatly increases the overall footprint of the computation. This can be partially mitigated by optimizing over $w_{add}$ (which we refer to as our \textit{add-warms} approach), since this allows the option to store magic states before the logical computation begins. It will be seen in Sec.~\ref{sec:final_estimates} that, while the latter approach does appreciably cut storage costs (therefore also reducing the storage contribution to total error, which has indirect consequences in the space-time volume through allowing lower code distances), it does not resolve the basic issue that keeping production constant but minimal throughout the computation causes the need to store large quantities of magic states due to irregularities in the consumption rate of real circuits. 

To resolve this issue, we have implemented a third \textit{min-storage} approach. In this approach, we force magic state production to merely keep up with magic state consumption by turning off factories wherever they would otherwise contribute to magic state excess. To accomplish this, we require there to be as many states in reserve at the beginning of any distillation cycle as there are requests within the next cycle. More formally, this means that $N(k) = m(k+1) - m(k)$, where $N(k)$ is the number of factories operating during the kth distillation cycle and $N(0) = m(1)$ is the number of factories operating during a single warm-up cycle. In this approach, the distillation contribution to active volume is modified according to $V_{dist} = \sum_{i=0}^{k_{max}-1} N(k) V(\mathcal{D})$ (note that in this scenario $k_{stop} = k_{max} - 1$). The formula (and associated approximation) used for $V_{storage}$ remains unchanged, though in this scenario $R(k) = N(k)$\footnote{It should be noted that in this setting the magic state storage costs become small enough that the costs of routing them between factories, storage banks, and their reserved tiles may no longer be negligible, though we still do not consider them here.}. While the min-storage approach leads to the minimization of storage costs, it causes there to be unused distillation tiles whenever $N(k) \neq N_{max}$ and a larger number of tiles dedicated to distillation than in other approaches since $N_{max}$ will match the number of states required by the highest-consumption distillation cycle. 

While we find (see Sec.~\ref{sec:final_estimates}) that the min-storage approach effectively resolves the storage issue, another way to mitigate the same problem while improving realism could be to adapt the default approach to a setting where a maximum size of quantum computer ($n_{total} d_2^2$) is specified. Given input values of $d_1$ and $d_2$ (which specify $n(\mathcal{D})$), this would effectively provide a coupling constraint between the number of factories ($N$) and $n_{storage}$. One could satisfy this constraint by varying a magic state storage capacity parameter $R_{max}$ to limit $n_{storage}$, assuming that all magic state factories remain on at all times and newly generated magic states replace old ones if the reserve is full. The number of magic state factories would need to increase (under the constraint of computer size) to meet production needs, since in this setting more magic states would be produced than consumed over the course of the computation. The constraint offered by setting a maximum computer size would, of course, not always be satisfiable. We do not explore the trade-offs introduced by this approach here. 

\section{Resource Estimates for Ground State Estimation Circuits}
\label{sec:final_estimates}

\subsection{Logical and Fault-Tolerant Resource Estimates}
In this section, we present resource estimates for large practical quantum circuits, namely those corresponding with one Trotter step of an efficient circuit implementation of the Trotter-Suzuki ground-state estimation (GSE) algorithm that was proposed in Ref.~\cite{kornellsome}. To compile this algorithm into the gate set detailed in Sec.~\ref{sec:compilation_strategy}, we have used the Quipper programming language~\cite{green2013quipper} together with LinguaQuanta~\cite{wesley2024linguaquanta}, a Quipper-to-QASM transpiler. We use Quipper's default precision of $10^{-30}$  when decomposing rotations into Clifford+T gates. From these circuits, we compile lattice surgery circuits and estimate resources for 5, 7, 9, 11, and 13 orbitals. See Table~\ref{tab:gse_circuits} for details on the numbers of logical qubits, logical gates, logical T/$\mathrm{T}^{\dagger}$ gates (equivalent to the number of MagicStateRequest LLI), compiled slices, sliced LLI processed by the LSC\footnote{The number of \textit{sliced} LLI differs from the number of LLI in that where an LLI appears on multiple slices (as is the case in BellBasedCNOT and RotateSingleCellPatch) it is multiply counted. This number does not include local instructions.}, and the LSC compile time. The largest circuit that we compile has around ten million gates and requires a little under fifteen hours to compile\footnote{While the compilation tasks were performed on relatively old (Skylake) CPUs available through the CADES cluster at ORNL, these compilation times highlight the need for advances in logical circuit decomposition to make practical quantum computing scalable, as mentioned in Sec.~\ref{sec:outlook}.}. Our resource analysis script, the functionality of which was detailed in Sec.~\ref{sec:resource_analysis}, uses one of three resource optimization approaches (`default', `add-warms', or `min-storage') to find the minimum value for an input cost function. We present results using all three of these approaches to optimize space-time volume, for which we use $n_{total}\times\tau_{total}\times d_2^3$ as a proxy. Our main results can be found in Tables~\ref{tab:gse_resources_default},~\ref{tab:gse_resources_add-warms}, and~\ref{tab:gse_resources_min-storage}, which show optimized parameter values ($d_1$, $d_2$, $N$, and $w_{total}$) as well as resource estimates: $n_{total}$ (in tiles), $\tau_{total}$ (in slices), $n_{total}\times \tau_{total}$ (in tile-slices), $V_{total}$ (in tile-slices), $n_{total}\times\tau_{total}\times d_2^3$ (our proxy for space-time), and $\epsilon_{total}$ (total error rate).

\begin{table*}[b]
  \centering
  \caption{Some details of the ground state estimation circuits that we estimate resources for.}
  \begin{tabular}{|c|c|c|c|c|c|c|}
    \hline
    \textbf{Orbitals} & \textbf{Logical Qubits} & \textbf{Gates} & \textbf{T/$\mathbf{T}^{\dagger}$ Gates} & \textbf{Slices} & \textbf{Sliced LLI} & \textbf{LSC Compile Time (s)} \\
    \hline
5 & 23 & 2.11e+05 & 8.31e+04 & 2.31e+05 & 2.70e+06 & 5.97e+02 \\
7 & 31 & 7.90e+05 & 3.10e+05 & 7.45e+05 & 1.00e+07 & 2.84e+03 \\
9 & 39 & 2.17e+06 & 8.47e+05 & 1.76e+06 & 2.74e+07 & 1.02e+04 \\
11 & 47 & 4.88e+06 & 1.91e+06 & 3.39e+06 & 6.16e+07 & 2.14e+04 \\
13 & 55 & 9.62e+06 & 3.75e+06 & 5.70e+06 & 1.21e+08 & 5.26e+04 \\
    \hline
  \end{tabular}
  \label{tab:gse_circuits}
\end{table*}

\begin{table*}[b]
  \centering
  \caption{Resource estimates and optimal parameter values for our default resource estimation approach (no additional warm-up cycles and the number of factories is constant).}
  \begin{tabular}{|c|c|c|c|c|c|c|c|c|c|c|}
    \hline
    \textbf{Orbitals} & \textbf{$d_1$} & \textbf{$d_2$} & $N$ & $w_{total}$ & \textbf{$n_{total}$} & \textbf{$\tau_{total}$} & \textbf{$n_{total} \times \tau_{total}$} & $V_{total}$ & \textbf{$n_{total} \times \tau_{total} \times d_2^3$} & $\epsilon_{total}$ \\
    \hline
5 & 5 & 19 & 11 & 3 & 3.59e+04 & 2.31e+05 & 8.27e+09 & 3.85e+09 & 5.68e+13 & 5.68e-03 \\
7 & 7 & 21 & 13 & 3 & 1.61e+05 & 7.46e+05 & 1.20e+11 & 5.62e+10 & 1.11e+15 & 4.27e-03 \\
9 & 7 & 23 & 15 & 3 & 4.51e+05 & 1.76e+06 & 7.95e+11 & 3.86e+11 & 9.68e+15 & 1.94e-03 \\
11 & 13 & 23 & 16 & 3 & 4.73e+05 & 3.39e+06 & 1.61e+12 & 7.37e+11 & 1.95e+16 & 3.67e-03 \\
13 & 11 & 25 & 17 & 3 & 9.65e+05 & 5.70e+06 & 5.50e+12 & 2.80e+12 & 8.60e+16 & 9.14e-04 \\
    \hline
  \end{tabular}
  \label{tab:gse_resources_default}
\end{table*}

\begin{table*}[htbp]
  \centering
  \caption{Resource estimates and optimal parameter values for our add-warms resource estimation approach (warm-up cycles are added and the number of factories is constant).}
  \begin{tabular}{|c|c|c|c|c|c|c|c|c|c|c|}
    \hline
    \textbf{Orbitals} & \textbf{$d_1$} & \textbf{$d_2$} & $N$ & $w_{total}$ & \textbf{$n_{total}$} & \textbf{$\tau_{total}$} & \textbf{$n_{total} \times \tau_{total}$} & $V_{total}$ & \textbf{$n_{total} \times \tau_{total} \times d_2^3$} & $\epsilon_{total}$ \\
    \hline
5 & 7 & 19 & 6 & 1786 & 1.13e+04 & 2.63e+05 & 2.97e+09 & 1.51e+09 & 2.03e+13 & 1.66e-03 \\
7 & 7 & 21 & 7 & 2931 & 4.07e+04 & 7.98e+05 & 3.25e+10 & 1.57e+10 & 3.01e+14 & 1.18e-03 \\
9 & 21 & 21 & 12 & 2864 & 8.71e+04 & 1.84e+06 & 1.60e+11 & 8.40e+10 & 1.48e+15 & 5.96e-03 \\
11 & 15 & 21 & 13 & 3942 & 6.18e+04 & 3.48e+06 & 2.15e+11 & 8.28e+10 & 1.99e+15 & 5.91e-03 \\
13 & 13 & 23 & 14 & 2314 & 9.80e+04 & 5.75e+06 & 5.63e+11 & 3.01e+11 & 6.85e+15 & 1.48e-03 \\
    \hline
  \end{tabular}
  \label{tab:gse_resources_add-warms}
\end{table*}

\begin{table*}[htbp]
  \centering
  \caption{Resource estimates and optimal parameter values for our min-storage resource estimation approach (factories are switched on to match consumption rate).}
  \begin{tabular}{|c|c|c|c|c|c|c|c|c|c|c|}
    \hline
    \textbf{Orbitals} & \textbf{$d_1$} & \textbf{$d_2$} & $N$ & $w_{total}$ & \textbf{$n_{total}$} & \textbf{$\tau_{total}$} & \textbf{$n_{total} \times \tau_{total}$} & $V_{total}$ & \textbf{$n_{total} \times \tau_{total} \times d_2^3$} & $\epsilon_{total}$ \\
    \hline
5 & 5 & 15 & 23 & 1 & 1.69e+03 & 2.31e+05 & 3.89e+08 & 7.46e+07 & 1.31e+12 & 5.99e-03 \\
7 & 5 & 17 & 33 & 1 & 1.94e+03 & 7.46e+05 & 1.45e+09 & 2.57e+08 & 7.11e+12 & 6.17e-03 \\
9 & 7 & 17 & 41 & 1 & 3.77e+03 & 1.76e+06 & 6.65e+09 & 9.37e+08 & 3.27e+13 & 3.84e-03 \\
11 & 7 & 17 & 47 & 1 & 4.28e+03 & 3.39e+06 & 1.45e+10 & 2.09e+09 & 7.14e+13 & 8.33e-03 \\
13 & 7 & 19 & 43 & 1 & 3.28e+03 & 5.70e+06 & 1.87e+10 & 3.66e+09 & 1.28e+14 & 1.16e-03 \\
    \hline
  \end{tabular}
  \label{tab:gse_resources_min-storage}
\end{table*}

\begin{figure}[htbp]
    \centering
    \includegraphics[width=0.9\linewidth]{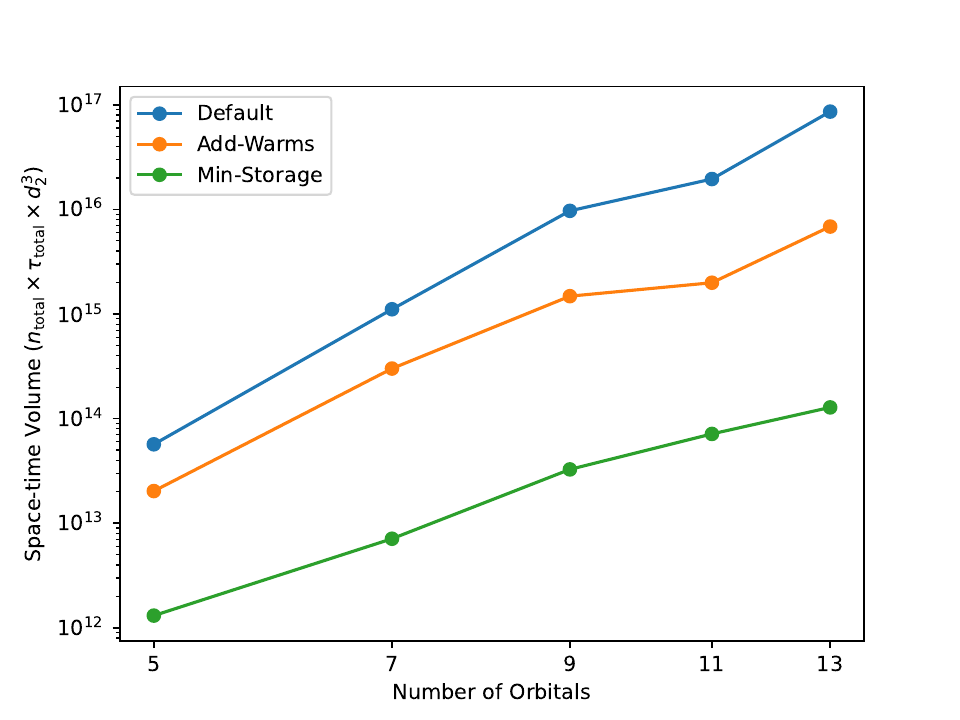}
    \caption{Comparison of space-time volume vs. number of orbitals between all three resource optimization approaches.}
    \hfill
    \label{fig:spacetime}
\end{figure}

\begin{figure}[t]
    \centering
    \includegraphics[width=1\textwidth]{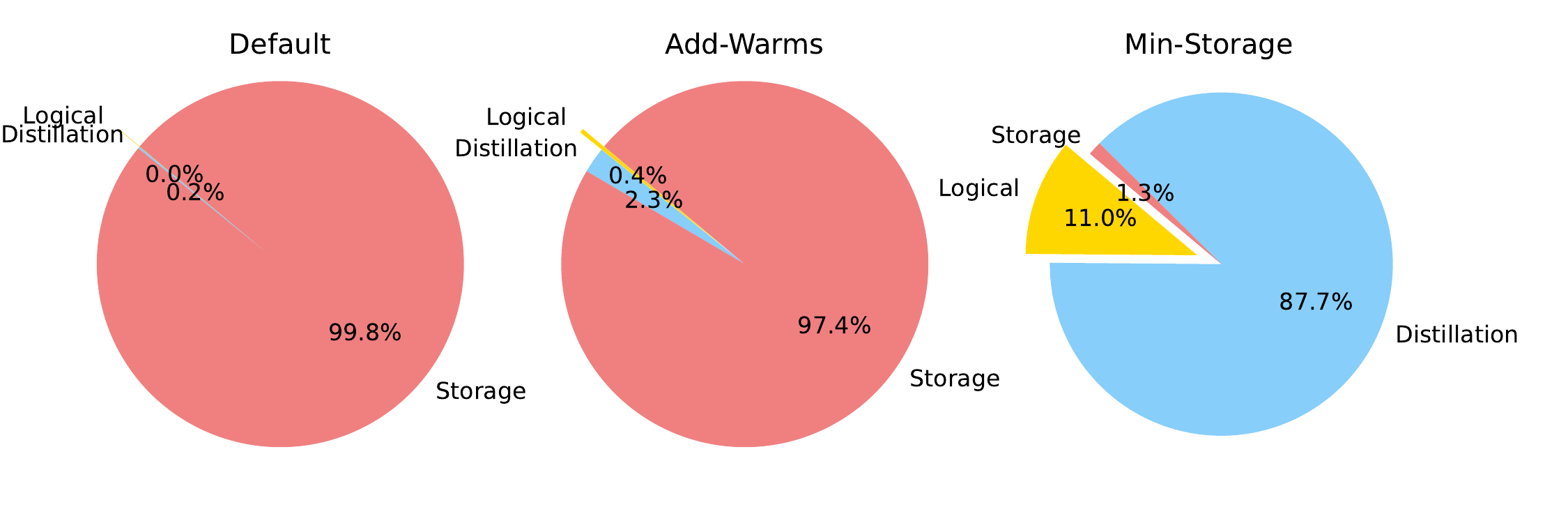}
    \captionsetup{skip=10pt}
    \caption{Proportion of quantum computer dedicated to the tasks of logical computation, magic state distillation, and magic state storage for the 13 orbital ground state estimation circuit using each of our resource optimization strategies (default, add-warms, and min-storage).}
    \label{fig:computer}
\end{figure}

First, we note that while we used $\epsilon_{thr} = 0.01$ (representing a success rate of 99\%) as a constraint to our optimization protocol, our results typically do not saturate this bound. Instead, as the number of orbitals (and thus the size of the circuit) increases and the code distances remain constant, the total error rate typically increases until the code distance needs to increment due to the total error passing $\epsilon_{thr}$, at which point $\epsilon_{total}$ drops\footnote{Deviations from this trend in our results seem to stem from the complexity that the two-level factory design adds to the trade space.}. Second, we notice that while $\tau_{total}$ values did not substantially change between resource optimization approaches (since the primary contribution to this quantity is the number of slices from the LSC output, which is not changing), the $n_{total}$ values change dramatically. This influences the most important result from Tables~\ref{tab:gse_resources_default},~\ref{tab:gse_resources_add-warms}, and~\ref{tab:gse_resources_min-storage}, which is the scaling of space-time volume with respect to the number of orbitals, shown in Fig~\ref{fig:spacetime}. The superiority of the min-storage approach over the default and add-warms approaches for optimizing space-time volume is evident. Since the primary reason that space-time is changing between approaches is that $n_{total}$ is changing, we now investigate how much of the quantum computer is being dedicated to each task (logical computation, magic state distillation, or magic state storage) between the three resource optimization approaches. As anticipated in Sec.~\ref{sec:resource_analysis}, we found that most of the spatial resources required by the computation in the default and add-warms approaches come from magic state storage. To see this most vividly, Fig.~\ref{fig:computer}(a) and (b) show that storage requires $\sim 99.8\%$ of the hardware tiles in the former case and $\sim 97.4\%$ in the latter, while it contributes the least to the overall spatial cost in the min-storage approach [Fig.~\ref{fig:computer}(c)]. In this latter case, magic state distillation becomes the dominant cost, bringing us into the regime that people typically consider when they neglect magic state storage. However, as mentioned in Sec.~\ref{sec:resource_analysis}, the min-storage approach involves allocating more space to distillation factories than is typically utilized within a given distillation cycle (the min-storage approach requires three times as many factories as the add-warms approach for 13 orbitals); therefore, our results show that minimizing resources dedicated to magic state distillation is not always beneficial, and that doing so while holding magic state production constant can be disastrous in real circuits with variable consumption rates due to the cost of storing excess states produced during lower-consumption sections of the circuit.

\begin{figure}[htbp]
    \centering
    \begin{subfigure}{0.49\textwidth}
        \centering
        \includegraphics[width=\linewidth]{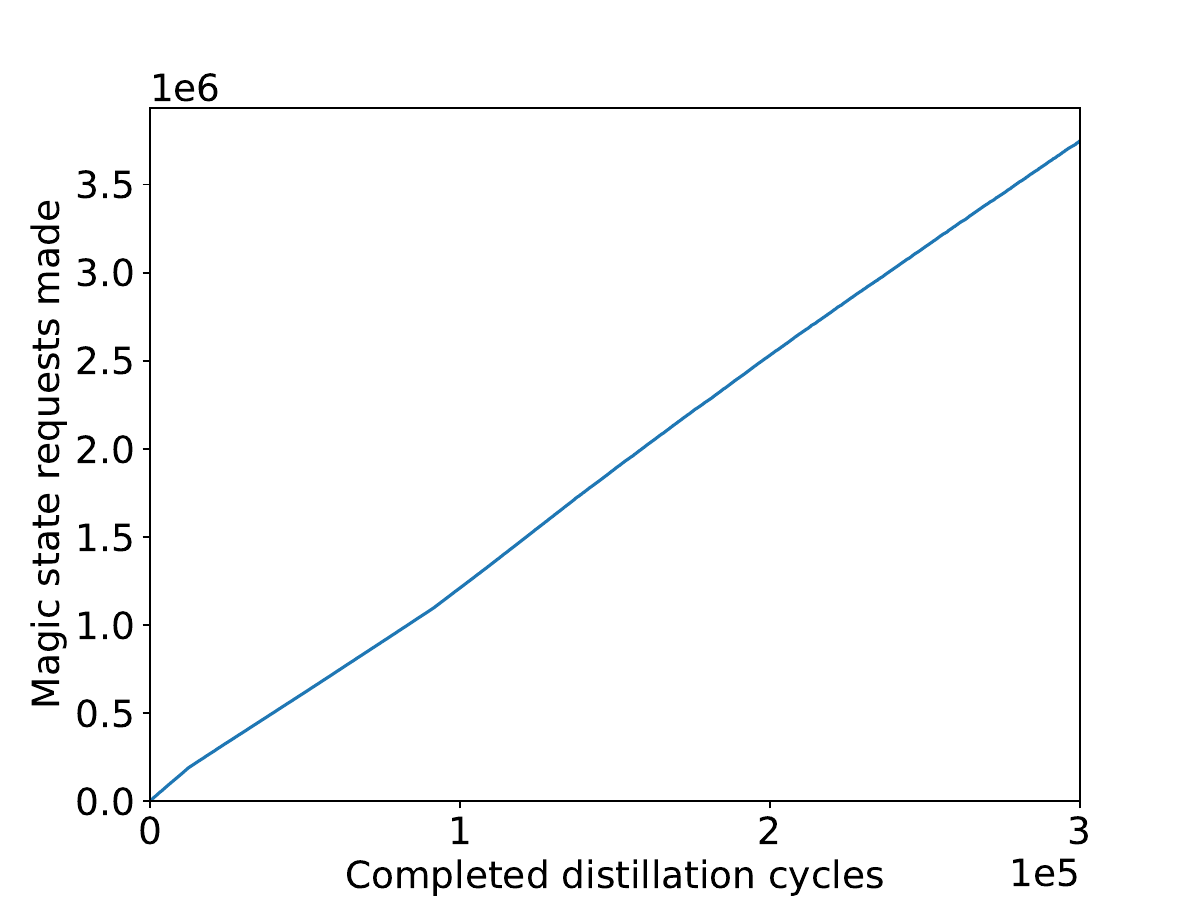}
        \caption{Total magic state requests made throughout circuit.}
    \end{subfigure}
    \begin{subfigure}{0.49\textwidth}
        \centering
        \includegraphics[width=\linewidth]{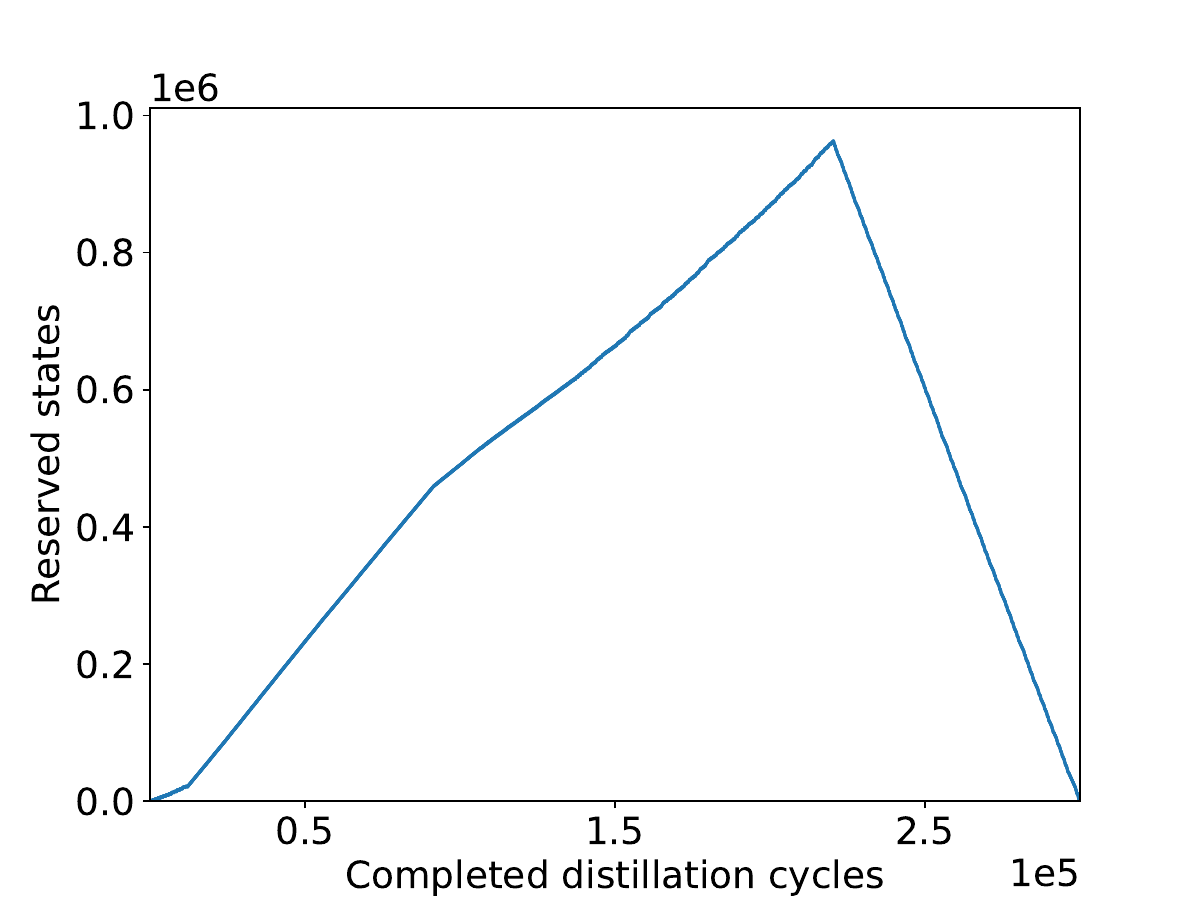}
        \caption{Number of states in reserve for default approach.}
    \end{subfigure}
    \begin{subfigure}{0.49\textwidth}
        \centering
         \includegraphics[width=\linewidth]{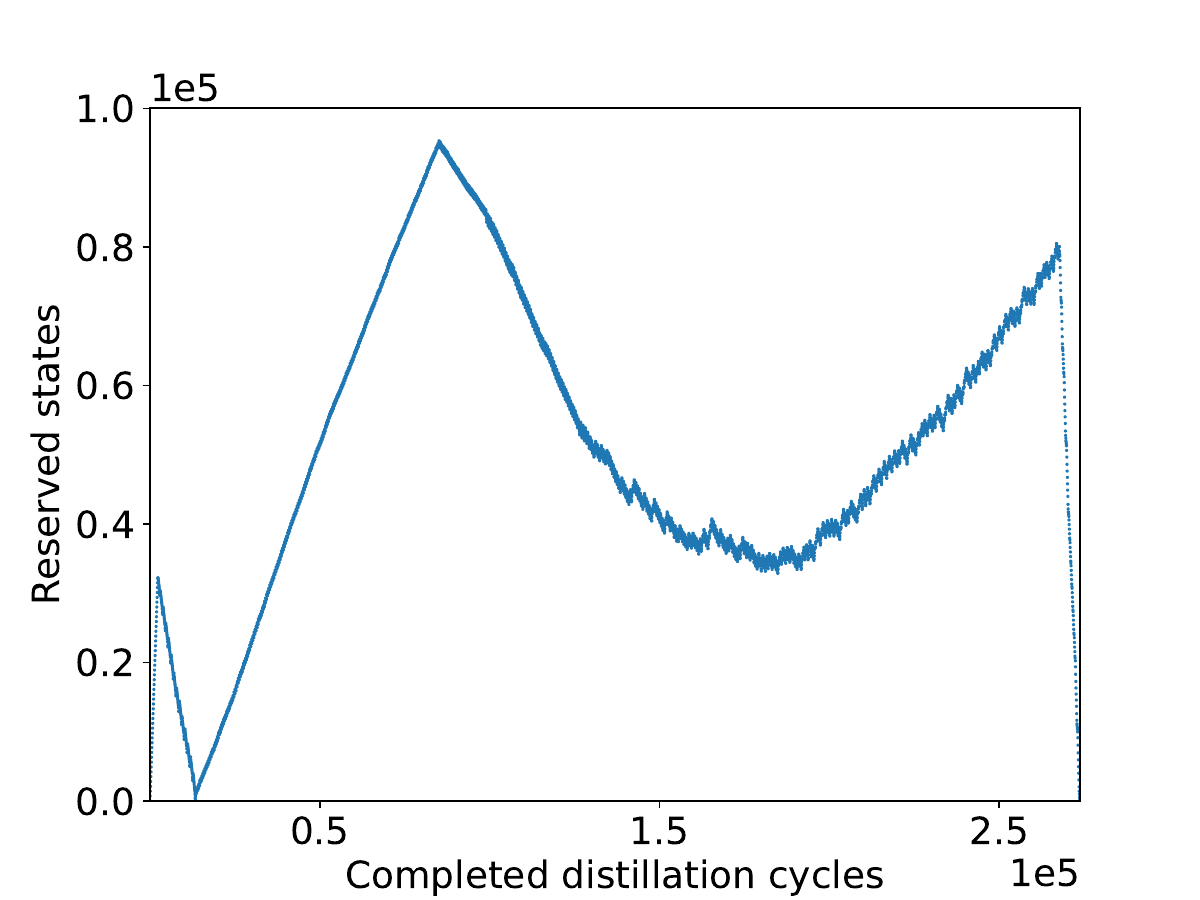}
        \caption{Number of states in reserve for add-warms approach.}
    \end{subfigure}
    \begin{subfigure}{0.49\textwidth}
        \centering        \includegraphics[width=\linewidth]{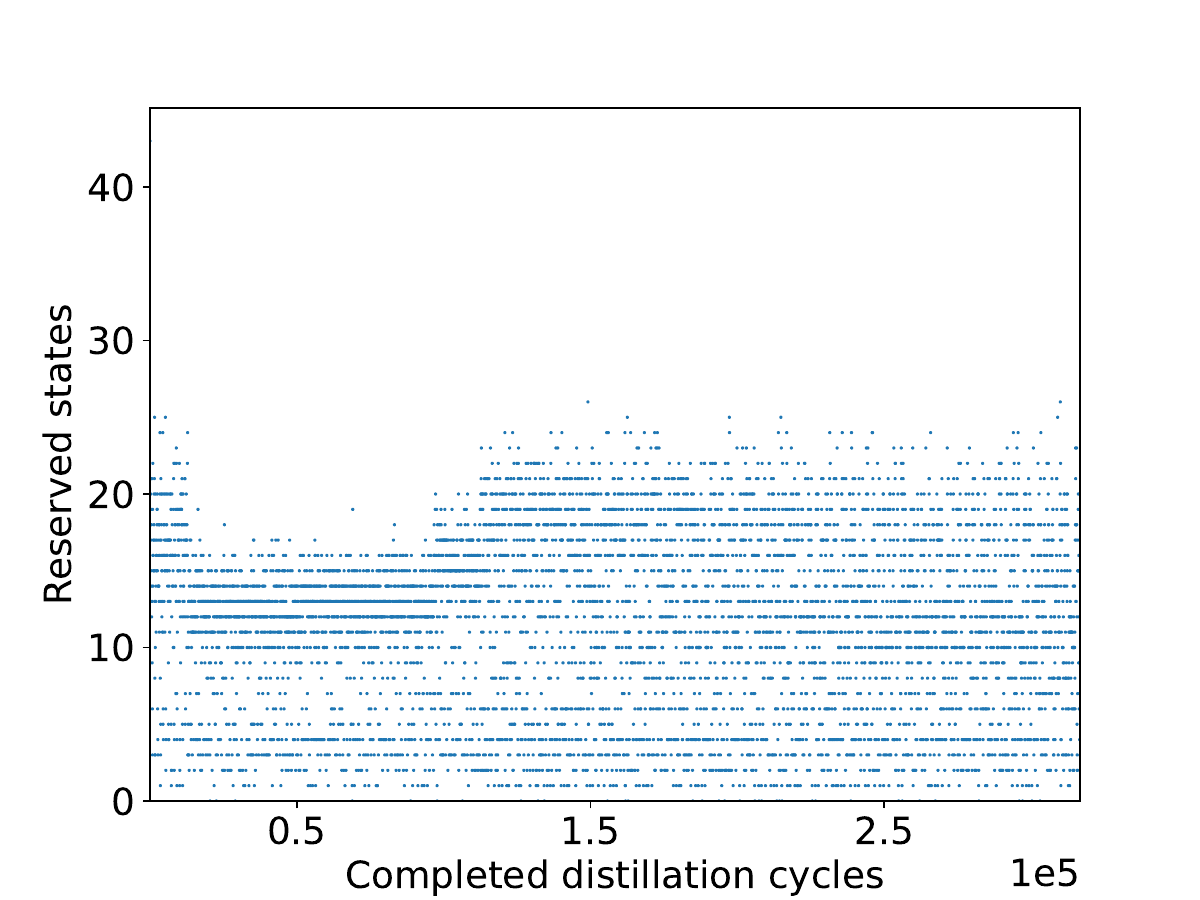}
        \caption{Number of states in reserve for min-storage approach.}
    \end{subfigure}
    \caption{Profile of total requested magic states coarse-grained by distillation cycle (a) and number of reserved magic states at the end of every fiftieth distillation cycle for three resource optimization approaches (b)-(d) for the 13-orbital ground state estimation circuit. Since the number of slices per distillation cycle differs by approach, the profile from the default approach is used in (a), though the profile is qualitatively similar in the other cases. }
    \label{fig:magicstaterequests}
\end{figure}

To make this last statement more concrete, Fig.~\ref{fig:magicstaterequests}(a) shows the cadence of magic state requests (coarse-grained by distillation cycle) for the 13-orbital GSE circuit in the default approach. Looking at the profile of magic state requests as a function of distillation cycle, we see that the rate of requests is slightly higher in the beginning of the circuit than in the bulk of the circuit; other irregularities in the rate are masked by the large number of requests being made. For comparison, Figs.~\ref{fig:magicstaterequests}(b)-(d) show the number of magic states in reserve as a function of distillation cycle for each of the three resource optimization approaches. Because our methodology ensures that, in all three approaches, magic state supply keeps up with demand at all times, the need for a higher production rate at the beginning can result in a growing excess of magic states during the bulk of the circuit until production halts once every needed magic state has been produced; that is, if the number of factories that are switched `on' is constant. This is especially apparent for the default approach in Fig.~\ref{fig:magicstaterequests}(b), but is also apparent for the add-warms approach in Fig.~\ref{fig:magicstaterequests}(c). In the latter case, adding warm-up cycles so that a larger bank of magic states exists at the beginning of the circuit does mitigate this problem to a great extent (the maximum number of states in reserve is about twenty times lower in this case), but it can not overcome the problem of having large numbers of excess magic states at various points in the circuit due to irregularities in consumption rate. In this case, a large bank produced at the beginning of the circuit, once depleted by the high-consumption portion of the circuit, gets re-built during lower-consumption portions until every needed magic state has been produced. Conversely, it can be seen for the min-storage approach in Fig.~\ref{fig:magicstaterequests}(d) that the number of states in reserve during any given distillation cycle is minimal, as expected by construction of that approach.

\subsection{Trapped-Ion Resource Estimates}

The resource estimates presented in the former section are in terms of hardware tiles and logical time-slices, which are abstract quantities from the fault-tolerant compilation layer. As of yet, our method has been entirely hardware agnostic except for the input of physical error rates (we have used the \verb+projected+ parameter set noted in Sec.~\ref{sec:resource_analysis})\footnote{While it is expected that fault-tolerance schemes will be co-designed with hardware architectures, we have opted for a general-purpose scheme.}. To obtain resource estimates in real terms (space and time) we will consider the cost to execute the same GSE circuit on a trapped-ion processor. To do this, we must consider the time to execute the slowest instruction from our local instruction set, as this will define the execution time for one slice, as well as the spatial requirement for a single hardware tile. To obtain these quantities, we make use of the Trapped-Ion Surface Code Compiler (TISCC) from Ref.~\cite{leblond2023tiscc}, which generates time-resolved circuits for the same set of local lattice surgery instructions considered in this paper. We will not consider details of the trapped-ion hardware model employed by TISCC here, and opt to simply state the final results. Using trapped-ion gate durations derived from the present literature, TISCC estimates a single round of error correction to require $9.613 \times 10^{-3}$ s. Our best resource estimates for the 13-orbital GSE circuit are from Table~\ref{tab:gse_resources_min-storage}, so we will use $d_2 = 19$, which yields the time cost of a logical time-slice to be $\tau_0 = 1.83 \times 10^{-1} s$\footnote{We did not consider decoding time to dominate the cost of a logical time-step because the rate of syndrome extraction for trapped-ion systems appears to be slower than state-of-the art decoding algorithms (see benchmarking for PyMatching in Ref.~\cite{Delfosse:2023cyx}).} . Querying TISCC further to estimate the spatial cost of a hardware tile at $d = 19$, we find $n_0 = 1.13 \times 10^{-3} m^2$. Combining these numbers with the results from Table~\ref{tab:gse_resources_min-storage}, we find that the total time to run the circuit is $t = \tau_{total}\times \tau_0 = 1.04\times 10^{6} s$, which is just over twelve days, and that the total area needed to support the hardware grid is $A = n_{total} \times n_0 = 3.71 m^2$. 

We do not believe that these space-time resource estimates are optimal, as we know of multiple ways in which they could be improved. Firstly, the ground state estimation circuit that has been used is an optimized implementation of the Trotter-Suzuki algorithm, but Trotter-Suzuki methods may not be optimal for ground state estimation. Secondly, our fault-tolerant scheme may not be optimal. As motivated in Sec.~\ref{sec:compilation_strategy}, there is a huge space of options to consider, and in this work we have opted to explore some of the trade-offs that exist within a single scheme. We expect that different approaches to translating circuits into lattice surgery primitives, such as more advanced optimization procedures using the ZX-calculus, could lead to much improved estimates. Thirdly, while we have used projected values for hardware gate infidelities in our analysis, TISCC uses estimates for gate times and trap widths that are based on current systems to perform its resource analysis. By the time that a trapped-ion quantum computer is able to run an algorithm fault-tolerantly using surface code lattice surgery, values for all of these parameters may significantly improve. It is also well-known that trapped-ion processors are slower than those of e.g. superconducting circuits, with syndrome extraction circuits expected to be $\propto 1000\times$ faster~\cite{beverland2022assessing} in the latter, though trapped-ion systems have longer coherence times. It is not clear how this trade-off impacts the space-time cost of running an algorithm fault-tolerantly.

\section{Conclusion and Outlook}
\label{sec:conclusion}

In this work, we have presented a resource estimation pipeline for practical quantum circuits that relies on explicit lattice surgery compilation to obtain an accurate number of logical time-steps, an accurate active volume, and a profile of fulfilled magic state requests by time-step, all of which significantly improve the realism of our estimates relative to existing approaches. In order to implement our methodology, we have heavily revised the Lattice Surgery Compiler~\cite{watkins2023high} to include modified instruction sets, a local compilation layer, improved layout generation, and DAG-based parallelism, among other things. We utilized three different resource optimization schemes within our pipeline to estimate resources for ground-state estimation circuits up to 13 orbitals to explore trade-offs related to magic state distillation and storage. We showed that assuming a constant (yet minimal, under the condition that there are always at least enough magic states in storage to fulfill all requests during a given distillation cycle) number of active magic state factories causes an unreasonable amount of space dedicated to storage in the quantum computer. Lastly, we showed how to combine our fault-tolerant resource estimates with results from a surface code compiler for trapped ion processors (TISCC)~\cite{leblond2023tiscc} to obtain resource estimates in terms of space and time.

The work presented here relies on an abstract model of a quantum computer and is therefore hardware-agnostic so long as the local, tile-based surface code instruction set that has been used is desirable, though we understand that other instruction sets may be preferable for some hardware architectures (see, for example, a surface code instruction set specific to neutral atom systems~\cite{viszlai2023architecture}). While in combination with Ref.~\cite{leblond2023tiscc}, the work presented here aids in the development of an end-to-end compilation and resource estimation pipeline for trapped-ion processors in which the final result has been verified to be correct using simulations of those systems, our pipeline was designed to be extensible to other approaches.

In this paper, we have used the simplification that $\epsilon_{logical} = V_{logical}P(d)$, i.e. a uniform logical error rate was used for the entire active volume regardless of what operation was being performed. To improve the analysis for trapped-ion systems specifically, TISCC-generated circuits could be used together with a realistic error model specific to trapped-ion systems to produce logical error rates for each member of the output instruction set. These results, which could readily be incorporated into the methodology presented in Sec.~\ref{sec:resource_analysis}, would significantly improve its realism. We reserve the computation of such error rates for a future study.

At this stage of development in fault-tolerant compilation and resource estimation, we do not expect any resource estimates to perfectly predict what a future system could achieve, as numbers can vary wildly according to different approaches; see for example, the trade-offs between magic state distillation and storage that we have considered in this work. Nonetheless, we believe our resource estimation pipeline achieves improved realism over other approaches through explicit lattice surgery compilation that accounts for circuit-level parallelism and the management of magic states. We anticipate that future work could better characterize the trade-space by further expanding the functionality of the LSC to include different compilation strategies and/or by considering integration with different hardware compilers like TISCC.

\begin{acks}
We would like to thank Peter Selinger for many great conversations, and for his support with the Quipper implementation of the GSE algorithms considered here. We would also like to thank Scott Wesley for his help with LinguaQuanta.

The work of T. LeBlond and R. Bennink was supported by the Defense Advanced Research Projects Agency (DARPA) Quantum Benchmarking (QB) and Underexplored Systems for Utility-Scale Quantum Computing (US2QC) programs under award numbers HR00112580540001 and HR0011261528. The work of C.~ Dean was supported by the DARPA QB program under award number HR001122C0066 and was supported by the Air Force Office of Scientific Research, Air Force Material Command, USAF under award number FA9550-21-1-0041. The work of G. Watkins was supported by the DARPA QB program under award numbers HR00112230006 and HR001121S0026 and was supported by the QuantERA grant EQUIP through the Academy of Finland, decision number 352188. The views, opinions and/or findings expressed are those of the author(s) and should not be interpreted as representing the official views or policies of the Department of Defense or the U.S. Government. 

This manuscript has been authored by UT-Battelle, LLC, under contract DE-AC05-00OR22725 with the US Department of Energy (DOE). The US government retains and the publisher, by accepting the article for publication, acknowledges that the US government retains a nonexclusive, paid-up, irrevocable, worldwide license to publish or reproduce the published form of this manuscript, or allow others to do so, for US government purposes. DOE will provide public access to these results of federally sponsored research in accordance with the DOE Public Access Plan (https://www.energy.gov/downloads/doe-public-access-plan).

This research used resources of the Compute and Data Environment for Science (CADES) at the Oak Ridge National Laboratory, which is supported by the Office of Science of the U.S. Department of Energy under Contract No. DE-AC05-00OR22725.
\end{acks}

\bibliographystyle{ACM-Reference-Format}
\bibliography{refs}

\appendix
\end{document}